\documentclass[12pt,preprint]{aastex}

\newcommand{\dg}{${}^\circ$}
\newcommand{\mn}{${}^{\prime}$}
\newcommand{\scn}{${}^{\prime\prime}$}

\newcommand{\hii}{H\thinspace II}

\usepackage{float}                                                                           

\begin{document}


\title{The Effect of Metallicity on Cepheid-Based Distances
\footnote{Based on observations made with the NASA/ESA Hubble
obtained at the Space Telescope Science Institute, which is operated by
the Association of Universities for Research in Astronomy,
Inc., under NASA contract NAS 5-26555. These
observations are associated with program GO-8584.}}

\author{Shoko Sakai}
\affil{Division of Astronomy and Astrophysics, University of California,
Los Angeles, \\
Los Angeles, CA, 90095-1562}

\author{Laura Ferrarese}
\affil{Rutgers University, Department of Physics and Astronomy,\\
136 Frelinghuysen Road,  Piscataway, NJ 08854}

\author{Robert C.\ Kennicutt, Jr.} 
\affil{Steward Observatory, University of Arizona, Tucson, AZ  85721}

\author{Abhijit Saha}
\affil{WIYN Telescope, National Optical Astronomy Observatories, Tucson, AZ, 85726}

\begin{abstract}

We have used the Wide Field and Planetary Camera~2
on board the Hubble Space Telescope to 
obtain $V$ and $I$ images of seven nearby galaxies. For each,
we have measured a distance using the tip of the red giant
branch (TRGB) method. By comparing the TRGB distances to
published Cepheid distances, we investigate the metallicity dependence of the
Cepheid period-luminosity relation. Our sample is supplemented by 10 additional
galaxies for which both TRGB and Cepheid distances are available in
the literature, thus providing a uniform 
coverage in Cepheid abundances between 1/20 and  2 (O/H)$_\odot$.
We find that the difference between Cepheid and TRGB distances 
decreases 
monotonically with increasing Cepheid abundance, consistent with a 
mean metallicity dependence of the Cepheid 
distance moduli 
of ${{\delta{(m - M)}}/{\delta[O/H]}} =  {-0.24 \pm 0.05}$ mag~dex$^{-1}$.  

\end{abstract}


\section{Introduction}

In the past decade, the uncertainty in the value of the Hubble
constant based on the local distance scale ladder has decreased from
roughly a factor of two to $\pm$10--15\% (e.g., Mould, Kennicutt,\&
Freedman 2000).  This breakthrough was made possible by the determination of
HST-based Cepheid distances to 25 nearby galaxies, carefully selected to
provide an accurate calibration for a variety of secondary distance indicators
(Kennicutt et al.\ 1995, Saha 1997).  The dominant 
source of systematic errors in the distance scale as a whole and in H$_0$ in particular
resides in the
calibration of the Cepheid period luminosity relation, most notably its zero point (which is tied
to the distance to the Large Magellanic Cloud), and possible dependence (both in zero point and slope) on the metallicity of the variable stars.
Reducing these extant  errors is imperative. For instance, the recent WMAP analysis of
fluctuations in the cosmic  microwave background has produced a
value of the Hubble constant $H_0 = 71 \pm 4$ 
km~s$^{-1}$~Mpc$^{-1}$ (Bennett et al.\ 2003). While in perfect agreement
with the local value derived by the HST Key Project on the Extragalactic
Distance Scale (72 $\pm$ 8 km~s$^{-1}$~Mpc$^{-1}$; Freedman et al.\ 2001, hereafter F01), tighter constraint on the latter would allow a more meaningful comparison of these two estimates. 

The uncertainty in the
metallicity dependence of the Cepheid period-luminosity (PL) relation
is particularly troublesome. The galaxies used by the Key Project span
a range of gas-phase metal abundances of roughly a factor of 50 
($-1.5 \le [O/H] \le 0.3$; Ferrarese et al.\ 2000a), wide enough that a systematic change of 
0.5 mag in the Cepheid distance moduli per factor 10 increase in abundance 
would by itself introduce a systematic
error of approximately 10\% in the Key Project value
of H$_0$ (Kennicutt et al.\ 1998, hereafter K98; 
Mould et al.\ 2000; F01). Not only there are significant
metallicity offsets between Cepheids in the Key Project  galaxies and those in the 
Large Magellanic Cloud (LMC), on which the PL calibration itself rests,  such
offsets are different for  the mean samples used to calibrate 
secondary distance indicators (e.g., SNe~Ia, fundamental plane). This can potentially lead to
systematic offsets in the values of $H_0$ derived from individual calibrators.
A lack of constraints on the metallicity dependence of the Cepheid PL relation also
hampers the interpretation of fully external
tests of the zero point of the Cepheid distance scale (e.g., Hernstein et al.\ 1999).

The magnitude of the metallicity dependence of the Cepheid PL relation is, unfortunately, 
poorly constrained,
both theoretically or observationally.  Following K98, we
describe such dependece in terms of the parameter $\gamma$:

\begin{equation}
\gamma = {\delta {{(m - M)}_0}} / {\delta{\log Z}},
\end{equation}

where $\delta{{(m-M)}_0} = (m-M)_{\mbox{0,Z}} - (m-M){\mbox{0,LMC}}$, the difference
between the distance modulus obtained with and without the metallicity
correction, and $\delta{\log Z} = (\log Z)_{\mbox{LMC}} - 
(\log Z)_{\mbox{galaxy}}$.
Note that $\gamma$ reflects the net effect on {\it distance determination}.                    
Metallicity can affect both the luminosity of a Cepheid, as well as the                        
color boundaries of the instability strip. Since a mean period-color relation                  
is used to deduce reddening and extinction, the second of these effects can be                 
the dominating influence. Thus $\gamma$ really depends on the specific                         
passbands chosen. In this paper we are mainly concerned with Cepheid distances                 
based on $V$ and $I$ observations ($\gamma$) which covers essentially all                 
of the HST measurements.                                                                       

Theoretical models of Cepheids predict metallicity dependences
ranging from near zero (Saio \& Gautschy 1998; Alibert et al.\ 1999) to
significant dependences (in either direction!) of up to 
$\pm$0.3 mag~dex$^{-1}$ (Chiosi, Wood, \& Capitanio
1993; Bono et al.\ 1999; Sandage, Bell, \& Tripicco 1999;
Caputo et al.\ 2000; Fiorentino et al.\ 2002).  
Recent empirical determinations of $\gamma$ have yielded an even
larger range of values.
The HST Key Project attempted to constrain $\gamma$
in two ways, by comparing measured PL relations for two Cepheid fields
in M101 differing in [O/H] by 0.7 dex, and by investigating a possible
systematic difference between Cepheid and tip of the red giant branch (TRGB)
distances for a sample of 10 galaxies (K98).  These two tests
yielded marginal (1.5 $\sigma$) detections of a metallicity dependence,
with $\gamma = -0.24 \pm 0.16$ and $-0.12 \pm 0.08$ mag~dex$^{-1}$
respectively. This led to a provisional
correction of $-0.20$ mag~dex$^{-1}$ to the final Cepheid distances
published by the Key Project team (F01).  However, 
value of $\gamma$ between 0 and  $-$0.9 mag dex$^{-1}$ are supported by independent studies.
By comparing Cepheid, TRGB, and
RR Lyrae distances to the Magellanic Clouds and IC~1613, Udalski et al.\ (2001) 
detected no significant metallicity dependence.  A null result was 
also derived by Ciardullo et al.\ (2002) from a comparison of Cepheid and
planetary nebula luminosity function (PNLF) distances to nearby galaxies.
On the other hand, a comparison of Cepheid PL relations in the LMC and SMC
by Sasselov
et al.\ (1997) produced $\gamma = -0.4{^{+0.1}_{-0.2}}$.  A similar
analysis, but applied to the Key Project galaxies, yielded
$\gamma = -0.4 \pm 0.2$  (Kochanek 1997).  An even larger dependence was reported by 
Gould (1994), based on a re-analysis
of M31 Cepheid observations of Freedman \& Madore (1990).  The reason for
the discrepancy between the various studies is the
small number of galaxies used, the limited 
range of metal abundances spanned, and/or lack of
quantifiable systematic errors.

Most recently, Tammann, Sandage \& Reindl (2003) examined 321
Cepheid variables in the Galaxy with good $B$, $V$, and $I$ photometry
by Berdnikov et al. (2000), and compared them to more than 1000 Cepheids in 
the LMC and SMC (Udalski et al. 1999b,c). They found that the Cepheid variables followed
different period-color relations; LMC Cepheids were bluer than the Galactic
ones, for example.
They suggested that the observed differences in three galaxies for Cepheids
with $\log P > 1.0$ were due to metallicity differences.
Kanbur et al. (2003) then studied Cepheids from the HST Key Project,
and also from the Sandage-Tammann-Saha sample.
They measured the distances to these galaxies using several different
PL relations calibrated using Galactic and LMC Cepheids, including
the new relation by Tammann et al. (2003).
Kanbur et al. (2003) found that 
the Tammann et al. Galactic calibration yielded the same distances as
the Udalski et al. (1999) LMC calibration, if the latter were
corrected for a metallicity effect of $\gamma = -0.2$ from Freedman et al. (2001).
This suggested that the Galactic and LMC Cepheids did indeed follow different PL
relations, and constrained the metallicity dependence to be $\gamma \sim -0.2$ mag
dex$^{-1}$.

The goal of this paper is to correct all of these shortcomings
and perform a more robust test of
the metallicity dependence of Cepheid PL relation. We will follow the same technique used 
by K98, which was based on a comparison of Cepheid and TRGB distances for galaxies spanning 
a wide range in Cepheid metallicity. Compared to K98, our study benefits  from an increased 
sample size and a wider and more uniform range in Cepheid  metallicity. To  the 10 galaxies 
analyzed by K98, we add seven new measurements, covering a range in 
Cepheid abundances of 0.05 $-$ 2 in $Z/Z_\odot$. A comparison of  TRGB and Cepheid distances 
provides an
especially powerful test for a metallicity dependence of the Cepheid PL relation.The method is 
transparent and robust: 
over the metallicity range spanned by our galaxies, the TRGB magnitude in the $I$-band is 
insensitive to the metal abundance and age of the stellar population 
(Da Costa \& Armandroff 1990; Lee, Freedman \& Madore 1993; Salaris \& Cassisi 1997;
Sakai 1999). Furthermore, the metallicities of the halo fields targeted by TRGB observations 
do not correlate with those of the disk Cepheids, therefore even a small metallicity dependence
of the TRGB would not introduce any systematic biases in our results.
Finally, we also make the implicit assumption that the Oxygen fraction with respect
to the total metal content is constant for all galaxies used in the application
presented in this paper.

The paper is organized as follows: in \S2, we discuss the observations and reduction of the 
HST/WFPC2 TRGB data obtained as part of this program for six galaxies (plus one downloaded from 
the public HST archive).
\S3 deals with the TRGB distances, including those that had been published
prior to this paper. Cepheid distances, all of which 
have been previously published, are discussed in 
\S4. The Cepheid abundances, which are derived from those of nearby \hii\ regions,
are presented in \S5.
Results and discussion are presented in \S6 and \S7 respectively.

\section{HST Observations}

The key improvement of our study over the similar analysis conducted by K98 is the addition of new 
TRGB measurements for seven galaxies with well-determined Cepheid
distances: IC~4182, NGC~300, NGC~3031 (M81), 
NGC~3351, NGC~3621, NGC~5253, and NGC~5457 (M101).
Six of our new TRGB distances (all except NGC~5253)
are based on deep F555W ($V$) and F814W ($I$) images obtained
with the Wide Field and Planetary Camera 2 (WFPC2) on HST during Cycles 9 and 10
(GO-8584).  NGC~5253 was observed as part of an independent project and the data 
downloaded from the public HST Archive.
The TRGB distance to NGC~5253 is based on F814W observations only, which are however 
sufficient for an accurate determination.  Although five of the galaxies had already been observed
with HST with the goal of measuring  Cepheid distances,
the existing data were unsuitable for TRGB observations: the Cepheid fields
were placed in the crowded star forming disk regions, while
TRGB observations must target less crowded, metal-poor
halo regions, to allow for an unambiguous detection of the Population~II 
red giant stars.  

We restricted our sample 
to galaxies with Cepheid distances of $\le$10 Mpc, to assure reliable
detection of the TRGB within reasonable integration times.
Four of the galaxies (M81, M101, NGC~3351, NGC~3621) contain Cepheid
fields with metallicities higher than the LMC ($Z > 0.4 Z_\odot$),
where the K98 results are particularly poorly constrained.  
The most distant of our galaxies, NGC~3351, is especially critical since it contains
one of the most metal-rich Cepheid fields in the Key Project sample
(2.2 $Z_\odot$).  Another key target is
M101, for which Cepheid were observed in two separate fields with mean abundances of 
$\sim$0.4$~Z_{\odot}$ and $2~Z_{\odot}$ (K98).

The observations are summarized in Table~1, and the 
the WFPC2 field of view is superimposed to a ground based image of each galaxy 
in Figure~1. The exposure times 
were chosen to reach at least one magnitude below the 
TRGB in both F555W and F814W.  


The reduction and stellar photometry of the images was carried out
independently using the DAOPHOT (Stetson 1994) and DoPHOT (Schechter, Mateo,
\& Saha 1993) families of PSF-fitting procedures.  The reduction
procedures closely followed the methods described in the HST Key Project
series, with the exception that here we only deal with single-epoch observations.  
We refer the reader to Ferrarese et al.\ (1996) for a detailed
discussion of the reduction procedures, and only briefly summarize
the process here.

For both photometric reductions, the WFPC2 images were 
first calibrated using a standard pipeline maintained by the Space Telescope Science
Institute (STScI), which included
correction of analog-to-digital (A/D) conversion errors, detector
bias, dark, and flatfield corrections.
In addition, bad pixels were masked using the data-quality files provided
by the HST data processing pipeline, 
the vignetted edges of the detectors were blocked by applying an image mask, and 
photometric variations introduced by the geometric distortion of the 
WFPC2 optics were corrected using pixel
area maps. The frames for individual exposures were then 
co-added to make deeper $F555W$ and $F814$ images; in the process cosmic-rays were identified 
and removed using standard IRAF/STSDAS routines\footnote{IRAF is distributed by the National 
Optical Astronomy
Observatories, which are operated by the Association of Universities for
Research in Astronomy, Inc., under cooperative agreement with the National
Science Foundation.}
Finally, each frame was multiplied by 4 and converted to short integers.

The reduced, combined images were
processed with the DAOPHOT and ALLSTAR software to extract PSF magnitudes
from each image. These were converted to 
the calibrated Landolt (1992) system as described in detail in 
Hill et al.\ (1998).
The instrumental PSF magnitudes were first transformed to 0\farcs5 diameter
aperture magnitudes as defined by Holtzman et al. (1995)
by selecting $\sim$15 bright, isolated stars on each WFPC2 chip.
All other stars were subtracted to provide clean sky measurements, and 
aperture magnitudes were measured at 12 radii between 
0\farcs15 and 0\farcs50 to define a growth curve.
The magnitudes were then transformed to calibrate $V$ and $I$ magnitudes as
described in Hill et al. (1998).


PSF fitting magnitudes were independently measured using a version of  DoPHOT developed specifically to handle the peculiarities of the HST
data and PSFs (see Saha et al. 1994).
DoPHOT output magnitudes are simply proportional to the height of
the fitted PSF, and must be transformed to 0\farcs5 magnitudes (as in Holtzman et al. 1995)
by applying an aperture correction. In NGC 3351 and NGC
3651, aperture corrections could be calculated reliably (with an rms
uncertainty of 0.02 mag or less) from bright, isolated stars in the
field. For all other galaxies, not enough isolated stars exist to
allow for a determination of the aperture corrections from the fields
themselves; in these cases we adopted `standard' aperture corrections
calculated from independent observations of uncrowded fields in Leo I (Hill et
al. 1998). The difference between the Leo I aperture corrections, and
those calculated from the NGC 3361 and NGC 3651 data is at most 0.02 mag, and
agree identically for most chips. The
magnitudes thus obtained were converted to the `ground system'
magnitudes F555W and F814W as defined in Holtzman et al. (1995b) using
the zero points derived from observations of $\omega$ Cen (Hill et al.
1998). Finally, F555W and F814W magnitudes were converted to $V$ and
$I$ magnitudes following the procedure outlined by Holtzman et al.
(1995b).


The photometric results of the DAOPHOT and DoPHOT analyses were compared;
for both filters, the magnitudes of individual stars agreed
within the photometric errors.  In order to simplify the presentation
of the results, we only show the DAOPHOT/ALLFRAME photometry
in the remainder of this paper.

We stress that all of the photometry presented in this  paper is based on the  calibration of Hill et al. (1998). Updated calibrations do exist, 
using improved  
charge-transfer-efficiency corrections (e.g., Stetson 1998; Dolphin 2000).
The Hill et al.\ calibration was adopted in this analysis
to maintain consistency with the large majority of the published 
Cepheid photometry, against which our TRGB distances will be compared.
If the $V$ and $I$ WFPC2 zeropoints of Stetson (1998) were adopted instead, the
TRGB distance moduli derived in this paper would decrease by 0.04 mag,
and the HST-derived Cepheid distance moduli by 0.07 mag.  As discussed in 
\S6, because of the differential nature of our test, this has no effect on the 
results derived in this paper.

\section{TRGB Distances}

A detailed review of the TRGB method can be found in
Madore, Freedman, \& Sakai (1997) and Sakai (1999).   
As the stars become brighter and evolve up the red giant branch, 
they undergo a drastic change at the onset of helium-burning in the core.
At most wavelengths, the absolute magnitude of the
RGB tip is sensitive to the age and metallicity of the red giant population.
However in the $I$-band the tip luminosity 
has been shown, both observationally
and theoretically, 
to vary very little with $M_{I,TRGB} = -4.0 \pm 0.1$ mag
for stellar population ages between 2 and 15 Gyr, and
metallicities spanned by Galactic globular clusters ($-2.2 \leq$
[Fe/H] $\leq -0.7$)
(Da Costa \& Armandroff 1990; Lee et al.\ 1993; Salaris \& Cassisi 1997; 
Sakai 1999) 
The near constancy of the $I$-band tip magnitude produces a sharp edge in the luminosity function (LF), 
making the TRGB a reliable distance indicator.
Furthermore, the TRGB is calibrated independently of the Cepheid distance scale. These properties make the TRGB method
an ideal choice for constraining the metallicity dependence of
the Cepheid PL relation.

For this purpose, we obtained deep images of halo fields in the $V$ and 
$I$ bands, and determined the TRGB magnitude using automated techniques,
as described below.  Although the RGB tip
magnitude is determined from the $I$-band LF, the 
$I, (V-I)$ color-magnitude diagram (CMD) allows us to exclude more easily stellar
contaminants (e.g., AGB stars) and restrict the tip determination 
to stars with colors appropriate to the metallicity range within which the
TRGB calibration is reliable.

\subsection{TRGB Detection Methods}

Very early applications of the TRGB method relied on a visual estimate
of the TRGB magnitude from a CMD.  
Lee et al.\ (1993) showed that the application of a Sobel edge-detection filter 
with kernel [-1,0,+1] to the binned LF histogram can provide
an efficient and objective determination of the TRGB position.
Sakai, Madore, \& Freedman (1996) modified this method for application to a 
smoothed, continuous LF.
First, the smoothed $I$-band LF is represented by 
replacing the discretely distributed stellar magnitudes with the corresponding gaussians:

\begin{equation}
\Phi (m) = \sum_{i-1}^{N} \frac{1}{\sqrt{2\pi} \sigma_i} \exp \left[-\frac{(m_i-m)^2}{2\sigma^2_i}\right],
\end{equation}

\noindent where $m_i$ and $\sigma_i$ are the magnitude and photometric error 
of $i$th star,
respectively, and $N$ is the total number of stars in the sample.
The edge-detection filter is then defined by:
\begin{equation}
E(m) = \Phi(m+\sigma_m) - \Phi(m-\sigma_m),
\end{equation}
where $\sigma_m$ is the mean photometric error for all star with
magnitudes between $m-0.05$ and $m+0.05$ mag. 

Since its first application in Sakai et al. (1996), the uncertainty 
in the TRGB magnitudes derived using the edge-detection method 
has been quoted simply as the FWHM of the peak profile.
This is undoubtedly an overestimate; for instance, the
the peak of the Gaussian can be measured with higher precision than given by
its FWHM.
In this paper, the errors are estimated using a boot-strap test.
The magnitude of each star  is
varied randomly following a Gaussian distribution with  
$\sigma$ given by the observational error.
A new luminosity function is constructed using these
randomly-displaced magnitudes, and the TRGB
magnitude is determined using the edge-detection method.
This routine is repeated 500 times for each galaxy, and
the standard deviation of the distribution of 500 TRGB magnitudes
is taken as the uncertainty in the original tip determination.

The edge-detection method is very effective for galaxies when the RGB tip is located
$>$2 mag above the magnitude limit of the photometry, because in such cases,
the CMD is  well
sampled in this region, and crowding and
incompleteness are not significant.  However, the method becomes
less precise if sampling statistics are poor, or  
incompleteness strongly affects the LF
within 1 mag of the RGB tip location.  In such cases, the
edge detector response becomes noisy even when the RGB tip
is clearly seen in the CMD.
Therefore, in addition to the edge-detection method described above, 
we have also applied a modification of the cross-correlation (``CC") technique introduced by M\'endez et al.\ (2002).
A template $I$-band LF, $y_0(M)$, was constructed using the
LFs observed for fiducial galaxies, as described below.
The template LF is then compared to the giant-branch LF, $y(m)$,  of a given galaxy. 
The template is shifted in magnitude by increments $a$ and its 
normalization, $n$, is varied until the best match is found. This corresponds to
the minimum of the function $\phi(a) = \sum{_{n_{min}}^{n_{max}}} \sum{_{m_{min}}^{m_{max}}} |{(y_0(M)}-y(m+a))/n|$, where the summation
limits $m_{min}$ and $m_{max}$ are chosen to extend from 1 mag brighter
than the RGB tip to 1 mag below the tip (or when incompleteness
sets in).  These limits are intially set using the best estimate for
the RGB tip magnitude as determined from the edge method or visual 
inspection of the CMD, and then adjusted iteratively until convergence.  
The limits in the normalization constant are also varied around the value estimated
initially as the best ``guess''.


The template LF was constructed from the observed CMDs of
the halo regions of 
the nearby irregular galaxies IC~1613 (Freedman 1988),
IC~4182 (this paper), and NGC~5253 (this paper).  
Each galaxy has a well-sampled
LF at the RGB tip, and reliable photometry extending over one mag
below the tip.  The LFs were cross-correlated 
with each other within $\pm$2 mag of the RGB tips, and the 
shifted, matched, and normalized LFs were then added to form the final template.  
The resulting function $y_0(M)$ is shown
in Figure~2, shifted in magnitude so that the
tip has $I=0$ mag. We also overplotted the shifted, normalized 
luminosity functions of IC~1613, IC~4182 and NGC~5253.


The uncertainty in the TRGB magnitude determined using the CC technique
was estimated using two independent methods.
In the first, the TRGB magnitude of each galaxy was calculated using four different
template luminosity functions: the one discussed above 
(the combined luminosity
function), and the IC~4182, IC~1613 and NGC~5253 luminosity functions alone. 
The rms in the mean of four estimates was taken as the uncertainty.
This gives a rough estimate,
albeit not a very accurate one, since the combined 
and individual templates are not independent.
The second method uses the same boot-strap test described above, 
with the CC method applied at each of 500 iterations 
(a similar method was used by Mendez et al. 2002).
The derived uncertainty accounts for  errors introduced by photometric uncertainties
affecting the stellar magnitudes, 
while our first estimate of the uncertainty  in the tip magnitude is sensitive to
systematics hidden in the choice of the template luminosity function.
The two uncertainties are summed in quadrature to obtain the formal
error in the measured TRGB magnitude.
Even so, it needs to be pointed out that this error is likely a lower 
limit to the true uncertainty, since it does not account for other source 
of systematics, most notably crowding. For most of our galaxies,
the CC method returns errors between 0.01 and
0.05 mag. This is not altogether surprising, since a 0.05 mag mismatch 
between the template and the LF under study is generally clearly noticeable 
from a simple visual inspection.

\subsection{Calibration of the TRGB}

The calibration of the TRGB is discussed extensively by several
authors (e.g., DaCosta \& Armandroff 1990; Lee et al.\ 1993;
Madore et al.\ 1997; Salaris \& Cassisi 1998; Bellazzini, \& Ferraro,
\& Pancino 2001).  The zeropoint rests on observations of Galactic
globular clusters with  $-2.2 < $[Fe/H] $< -0.7$ 
(DaCosta \& Armandroff 1990; Lee et al.\ 1993).  The distances to
these clusters rest in turn on an adopted metallicity-$M_V$ calibration
for RR Lyrae stars, based on 
theoretical models of the horizontal branch for $Y_{MS}=0.23$ 
(Lee, Demarque \& Zinn 1990).  For this study we have adopted the
calibration of Lee et al.\ (1993), which is expressed as
$(m-M)_I = I_{TRGB} - M_{bol} + BC_I$.
The bolometric magnitude, $M_{bol}$ and the bolometric correction,
$BC_I$, can be related to the metallicity of the RGB stars:
$M_{bol} = -0.19$[Fe/H]$-3.81$ mag, and $BC_I = 0.81 - 0.243 (V-I)_{TRGB}$.
The metallicity, [Fe/H], is expressed in terms of the color of the RGB 
stars 0.5 magnitude fainter than the TRGB:
[Fe/H]$=-12.65 + 12.6 (V-I)_{-3.5} - 3.3 (V-I)^2_{-3.5}$.
Combining these gives:
\begin{equation}
M{_I^{TRGB}} = 1.594 - 2.394 (V-I)_{-3.5} + 0.627 (V-I)^2_{-3.5} + 0.243(V-I)_{TRGB}.
\end{equation}

It should be noted that the above TRGB calibration is semi-empirical,
based on an RR Lyrae distance scale, combined with the theoretical
models of the horizontal branch stars.
Cassisi \& Salaris (1997) presented a purely theoretical calibration
of the TRGB magnitude by examining their theoretical stellar evolution
models (Salaris \& Cassisi 1997).
They reported that the semi-empirical TRGB calibration is
too faint by about 0.1 mag compared to the theoretical model. 
This, they suggest, is due to the poor sampling of RGB stars
in the Galactic globular clusters observed by Frogel et al. (1983),
which were used in the empirical TRGB calibration.
We emphasize that a precise calibration of the TRGB method is
not necessary for the purpose of this paper.
Since we are testing a {\it differential effect}, 
as long as the same calibration is applied to all the TRGB distances,
our result will not depend on the zero point of the TRGB calibration itself.
We are using the Lee et al. (1993) TRGB calibration for now, but
in Section~7, we discuss the effects of using another calibration.


\subsection{TRGB Distances to Individual Galaxies}

The TRGB distance moduli used in this paper are summarized in Table~2.
In this section, we describe in detail how each TRGB distance
was derived using the edge-detection and CC methods.



\subsubsection{IC~4182}

IC~4182 is an SA(s)m galaxy
with a Cepheid distance of 4.5 Mpc (Saha et al.\ 1994; F01).
The placement of the WFPC2 field of view is shown in Figure~1.
Our derived CMDs are shown in 
Figure~3, for all four WFPC2 chips (top) and
separately for the WF~2 and WF~3 chips (bottom), which are more representative of the 
galaxy's halo population (see Figure~1). 
The TRGB is seen clearly just below $I \sim 24$ mag, especially 
in the bottom panel.  On the right side of Figure~3,
we show the $I$-band LF, constructed from stars
on the WF~2 and WF~3 chips with $0.5 \le (V-I) \le 1.9$.  The edge detector
response functions, also drawn in the Figure, shows a 
firm detection of the TRGB at $I = 24.20 \pm 0.07$ mag.  We performed
the same fit to the logarithmic LF (bottom right panel
of Figure~3) with identical results.

The CC method was also applied.
Figure~11 shows
the template LF (dotted line)
shifted to provide the best match to the LF of the galaxy under study.
In the case of IC 4182, the CC method yields a TRGB magnitude of $I = 24.20 \pm 0.05$ mag.
The agreement between the edge-detection and CC results is not altogether surprising, 
since IC~4182 was used to build the cross-correlation template.

The foreground extinction along the line of sight to IC~4182 is 
$A_B = 0.059$ mag (Schlegel, Finkbeiner, \& Davis 1998), corresponding to
$A_I = 0.027$ mag (using the reddening law of Cardelli et al.\ (1989),
with $R_V = 3.1$).
Adopting the Lee et al.\ (1993) calibration, the TRGB magnitude is expected at
$M_I^{TRGB} = 4.08 \pm 0.05$ mag for $(V-I)_{TRGB} = 1.49$ mag and
$(V-I)_{-3.5} = 1.45$ mag.  

We have estimated the TRGB magnitude of IC~4182 using three methods.
Before determining the distance modulus, we need to find out how the
three estimates are related.
In order to do this,
we used the boot-strap method of estimating uncertainties in the TRGB magnitudes,
as described in the previous Section.
For each of the 5000 iterations, the TRGB magnitudes were measured using all three methods.
The first two methods which use edge-detection filtering are found to be
correlated with each other, almost one-to-one.  
In contrast, the CC method is very stable from one iteration to another, and
does not correlate with the first two methods at all.
Thus, the calculation of the final average magnitude was done in two steps.
First, the average value of the  two edge-filtering methods was estimated,
by deriving the rms 
in the mean of the distribution of all TRGB magnitudes found (for both
methods) for 5000 iterations.  
Finally, the average of all three estimates was measured by taking
the weighted mean of the result of the CC method and the average value
estimated for the two edge-filtering methods, since the CC method is
not correlated with the edge-filtering method.

We adopt a TRGB distance modulus for 
IC~4182 of $(m-M)_0 = 28.25 \pm 0.06$ mag, corresponding
to a linear distance of $4.5 \pm 0.1$ Mpc. Since the edge detection
and CC methods are not fully independent, we have used
a conservative estimate of the uncertainty.  

\subsubsection{NGC~5253}


NGC~5253 is a peculiar Im galaxy in the M83 group, with a Cepheid
distance of 3.4 Mpc (F01; also see \S 4).
This galaxy was not included in our HST Cycle 9 observing program, but 
archival HST observations exist which enabled us to measure its TRGB distance.
The WFPC2 field is shown in Figure~\ref{figure:n5253wfpc}. 
The CMD is shown in Figure~\ref{figure:n5253tip} for the
entire region covered by the WFPC2 (top left panel), 
and for stars in the halo region outside the
ellipse in Figure~\ref{figure:n5253wfpc} (lower left panel).

The $I$-band LFs (linear and logarithmic) and corresponding filter
responses are shown on the right side of Figure~\ref{figure:n5253tip}.
The $V$-band observations of NGC 5253 used the
F547M filter rather than the broader F555W filter. This is not a major concern, 
since only calibrated $I$-band 
magnitudes are required for TRGB distance measurements. 
The TRGB is detected at $I = 24.03 \pm 0.06$ and $I = 23.98 \pm 0.05$
mag in the linear and logarithmic filter responses, respectively.
Applying the CC method yielded a TRGB magnitude $I = 23.95 \pm 0.05$
(Figure~\ref{figure:ccresults}), which agrees well with the result obtained using the
logarithmic luminosity function edge-detection method.
Both are less susceptible to
small noise which affects the analysis of the linear luminosity function, and might 
therefore
provide a better estimate of the tip magnitude. Nevertheless, we adopt the average 
of the three determination as our best estimate of the tip magnitude. The $I$-band 
foreground extinction in the direction of NGC~5253 is 0.109 mag.
Since there are no $V$-band observations of NGC~5253,
we cannot estimate the $(V-I)$ color of the RGB stars 
which is necessary to determine the absolute magnitude of
the TRGB.  Instead, we adopt $-4.0 \pm 0.1$ mag for
the TRGB magnitude, as most of the magnitudes fall within 
the magnitude range of $-3.9$ and $-4.1$ mag.
Our final distance
modulus to NGC 5253 from TRGB is thus $(m-M)_0 = 27.88 \pm 0.11$ mag, 
corresponding to a linear distance of 3.6 $\pm$ 0.2 Mpc.




\subsubsection{NGC~300}

NGC~300 is an SA(s)d galaxy in the Sculptor group with a Cepheid distance
of 2.0 Mpc (Freedman et al.\ 1992, F01).
The detection of the TRGB in this galaxy is somewhat more difficult 
than the previous cases, because the WFPC2 field does not sample a 
large enough region to include large numbers of halo giants.  


We were able to detect the tip in the logarithmic LF only 
by binning the data so that  $\sigma_i$ in Equation~(2) is 
doubled,
but we failed to detect
any significant edge in the noisier linear luminosity function.
Figure~\ref{figure:n300tip} shows the CMDs, linear and logarithmic luminosity
functions, and corresponding 
edge-detection filter response functions.  The LF was constructed using
only stars with $0.5 \leq (V-I) \leq 2.0$ mag.  The TRGB can be seen in the CMD near 
$I \sim$ 22.6.
In the logarithmic luminosity function, the TRGB was detected at
$I=22.62 \pm 0.07$ mag, as seen clearly in the right bottom
panel in Figure~\ref{figure:n300tip}.


Because the luminosity function rises very gradually for more than two magnitudes,
the CC method breaks down for NGC~300 as there is no unique edge to search for.
Thus, we adopt the result of the edge-detection method applied to the logarithmic luminosity
function as our estimate of the tip magnitude.
Using the Lee et al.\ (1993) calibration, the absolute magnitude of the TRGB 
is predicted to be $M_{\mbox{TRGB}} = 4.05 \pm 0.05$ mag for
$(V-I)_{-4.0} = 1.5$ mag and $(V-I)_{-3.5} = 1.4$ mag.
The foreground extinction to NGC~300 is $A_I = 0.025$ mag.
Therefore, the final distance modulus of NGC~300 from TRGB is
$(m-M)_0 = 26.65 \pm 0.09$, corresponding to a distance of $2.14 \pm 0.09$ Mpc.


\subsubsection{NGC~3031}

NGC~3031 (M81) is an SA(s)b galaxy 
with a Cepheid distance of approximately 3.6 Mpc (Freedman et al.\ 1994, F01).
The CMD is shown in the 
left panel of Figure~\ref{figure:n3031tip}.
All stars are included in the top left panel, while only those found on chips
WF~3 and WF~4 (the outermost regions) are shown in the bottom left panel.
As was the case for IC 4182, the outer fields suffers less contamination from younger disk stars,
and were therefore used for the TRGB distance determination.

Stars with 
colors $(V-I) \le 1.4$ and $(V-I) \ge 2.5$ were excluded in constructing the LF shown on the right side of Figure~\ref{figure:n3031tip}.
The edge-detection method applied to the linear and logarithmic LFs give TRGB magnitudes of
$I = 24.02 \pm 0.09$ and $24.08 \pm 0.04$ mag respectively, as shown in the bottom right panels of Figure~\ref{figure:n3031tip},
If stars detected only in the F814W frames (which reach deeper magnitudes than the F555W frames) are included in the analysis (to limit the effect of incompleteness near the TRGB), the tip is detected at $I = 24.03 \pm 0.10$ mag.

The
CC method applied to the data yields a considerably fainter tip magnitude, $I = 24.34 \pm 0.15$.
This is due to the fact that 
the luminosity function rises very slowly over $\sim 0.8$ mag; the CC method then triggers 
on the mid-point of this rising ``edge'', which is significantly fainter than the true TRGB magnitude.


As our final determination of the tip magnitude we therefore adopt the average of the three 
separate TRGB determinations using the edge-detection
method, $I_{TRGB} = 24.13 \pm 0.06$ mag. The RGB 
colors $(V-I)_{\mbox{\small TRGB}} = 2.1$ mag and $(V-I)_{-3.5} = 1.8$ mag imply
a TRGB luminosity  $M^I_{\mbox{\small TRGB}} = -4.05 \pm 0.10$ mag.  
The foreground extinction along the line of sight to NGC~3031 is
$A_I = 0.155$ mag.
The final distance modulus to NGC 3031 from TRGB is thus
$(m-M)_0 = 28.03 \pm 0.12$ mag, corresponding to a linear distance of $4.0 \pm 0.2$ Mpc .

\subsubsection{NGC~3351} 

NGC~3351 is an SB(r)b galaxy in the Leo group, and the most distant galaxy
in this study.  Its radial velocity is 774 km s$^{-1}$, and its Cepheid
distance is 9.3 -- 10 Mpc (Graham et al.\ 1997; F01).
The galaxy lies near the WFPC2 limit for detecting the TRGB
within reasonable exposure times: 14 of our allocated 28
orbits were spent on this galaxy alone.

Figure~\ref{figure:n3351tip} shows the CMD of stars detected on 
the WF2, WF3, and WF4 chips (top left), and for stars located in the 
outer halo region shown by the triangular outline in Figure~\ref{figure:footprints}.  
From the CMD, the TRGB
appears around $I \sim 26.5 \pm 0.5$ mag.
The linear edge filter, applied only to the stars in the outer region, shows a strong peak 
at $I = 26.55 \pm 0.13$ mag;
however magnitude incompleteness becomes severe at approximately the same
magnitude, potentially biasing the filter response.  
We have also applied the filters to the stellar sample with blue stars excluded.
However, this exercise did not yield any result that is more accurate than 
using a larger sample, likely due to a very small number statistics.

The TRGB is hardly visible in the logarithmic LF, making for a very uncertain tip determination.
The CC method seems more robust, it provides a perfect fit to the rising part of the LF (see Figure 11), 
and yields a tip detection of $I = 26.53 \pm 0.10$ mag.





Taking the average of two TRGB magnitude estimates, we have
$I = 26.54 \pm 0.08$ mag.
The foreground extinction in the direction of NGC~3351 is $A_I = 0.054$ mag.
For $(V-I)_{TRGB} = 1.2$ mag and $(V-I)_{-3.5} = 1.1$ mag,
the predicted TRGB absolute magnitude is  $M{^I_{TRGB}} = 3.9 \pm 0.1$ mag.
Thus, the final distance moduli to NGC~3351 from TRGB
is $(m-M)_0 = 30.39 \pm 0.13$ mag, corresponding to a linear distance of
$12.0 \pm 0.7$ Mpc.

\subsubsection{NGC~3621}

NGC~3621 is an SA(s)d galaxy with a Cepheid distance of 6.6 Mpc
(Rawson et al.\ 1997; F01).
CMDs are shown in Figure~\ref{figure:n3621tip}; the 
halo stellar population is best represented in the WF2  chip (Figure~\ref{figure:footprints}).
On the top right panel of Figure~\ref{figure:n3621tip},
the I-band luminosity function of stars on WF2 is shown, together with
the corresponding edge filtering response function. 
No color cut to the luminosity function was applied as it did not make
the TRGB detection any better.
The TRGB is detected at $25.42 \pm 0.06$ mag.
Application of the edge-filtering to the logarithmic luminosity function 
yields $I = 25.47 \pm 0.06$ mag, while the CC method produces $I = 25.46 \pm 0.05$ mag, both consistent with the edge-filtering determination made using the linear LF.
We adopt the average of the three estimates as our final tip magnitude. 




The foreground extinction to NGC~3621
is $A_I = 0.156$ mag, and the observed color of the giant branch
implies a TRGB luminosity $M{^I_{TRGB}} = -4.06 \pm 0.10$ mag.
The final distance modulus to NGC 3621 from TRGB is thus $(m-M)_0 = 29.36 \pm 0.11$ mag,
corresponding to a linear distance of $7.4 \pm 0.4$ Mpc.


\subsubsection{NGC~5457}

NGC~5457 (M101) is an SAB(rs)cd galaxy with a Cepheid distance 
of $\sim$7 Mpc (Kelson et al.\ 1996; F01).
This galaxy is of particular interest for this project, since
two independent Cepheid distances have been estimated in two separate fields with different
mean metal abundances.

In Figure~\ref{figure:n5457tip}, we show CMDs for the entire WFPC2 field
(top left) and the outermost chips WF3 and WF4 (bottom left).
Only the latter were used when constructing the $I$-band LFs,
as shown on the right side of Figure~\ref{figure:n5457tip}.
The TRGB is clearly detected in the linear and logarithmic filter
responses, at $I = 25.41 \pm 0.04$ and $25.40 \pm 0.04$ mag respectively.
This field has a strong contamination of brighter red stars, presumably
AGB stars and red supergiants; this is not particularly surprising
because M101 has a very extended disk, making it difficult 
to isolate a purely halo-dominated field.  Despite this contamination
the RGB tip stands out clearly in the LFs and the filter responses.
We have also applied the edge-detection method to a sample with
the blue stars ($(V-I) < 0.5$ mag) excluded, and obtained
the exactly same result.
Applying the CC method gives best fitting tip magnitude of
$I = 25.42 \pm 0.05$ mag, which agrees very well with the edge-filter results.

The foreground extinction to NGC~5457 is $A_I = 0.017$ mag.
The TRGB calibration for this galaxy, with $(V-I)_{TRGB}=1.50$mag
and $(V-I)_{-3.5} = 1.36$ mag, is $M{^I_{TRGB}} = -4.02 \pm 0.10$ mag.
Taking the average of three TRGB magnitude estimates (two by
the edge-filtering method and one by the CC method),
we obtain $I_{TRGB} = 25.40 \pm 0.04$ mag.
Thus the distance modulus of NGC~5457 from TRGB 
is $29.42 \pm 0.11$ mag, corresponding to a linear distance of $7.7 \pm 0.4$ Mpc.

\subsection{Other TRGB Distances from the Literature}

In addition to the galaxies discussed in the previous sections, Table~2 lists published 
TRGB distance moduli for 10 additional galaxies with 
well determined Cepheid distances.  We comment briefly on the individual galaxies below.
In most case, we make use of color and extinction data data tabulated by Ferrarese et al. 
(2000a: F00) in converting tip magnitudes to distance moduli. 

LMC:  
The first estimate of the TRGB magnitudes, $I_{0,TRGB} = 14.53 \pm 0.05$ mag, was by 
Reid, Mould, \& Thompson (1987). They used data from the Shapley III star forming region, 
which is heavily contaminated by intermediate-age AGB stars.
A second estimate, $I_{0,TRGB} = 14.50 \pm 0.25$ mag, came from Romaniello et al. (2000) 
analysis of HST/WFPC2 observations of regions around SN1987A.  Unfortunately, the small 
spatial coverage of the WFPC2 field of view allowed them to detect only $\sim 150$ stars 
in the brightness range necessary for the TRGB measurement.
Cioni et al.\,(2000) also attempted to the TRGB distance based on a very large
stellar sample from the DENIS survey.
However, they were not able to constrain the internal reddening well enough
to estimate the accurate TRGB distance.
The most reliable determination is from Sakai, Zaritsky, \& Kennicutt (2000) who used 
data from the Magellanic Clouds
Photometric Survey (Zaritsky, Harris, \& Thompson 1997).
The unique feature of this study is that the reddening was determined
along the light of sight of individual stars by fitting spectra and
extinction to  $UBVI$ photometry (Zaritsky 1999).  
Therefore, Sakai et al. (2000) were able to select the regions
of low reddening, and reported $I_{0,TRGB} = 14.54 \pm 0.04$ mag.
Using $(V-I)_{TRGB} = 1.7 \pm 0.1$ mag, and $(V-I)_{-3.5} = 1.5 \pm 0.1$ mag,
the absolute TRGB magnitude is expected at $M_{I,TRGB} = -4.05 \pm 0.06$ mag.
We thus adopt $(m-M)_0 = 18.59 \pm 0.09$ mag as the LMC TRGB distance modulus.


SMC:  The TRGB distance to the SMC has been measured using
DENIS data by Cioni et al.\,(2000), who derived a modulus 
$(m - M)_0 = 18.99 \pm 0.03 \pm0.08$ mag.  

Sextans A:   Sakai, Madore \& Freedman (1996) detected 
the TRGB at $I=21.73 \pm 0.09$ mag using single-epoch ground-based data.  
More recently, Dolphin et al. (2003) used HST/WFPC2 observations to
detect the tip at $I = 21.76 \pm 0.05$ mag.
Since the HST/WFPC2 sample is very sparse (the I-band luminosity
function jumps by only a few stars at the TRGB edge),
we adopt the TRGB distance modulus from
Sakai et al. (1996), $(m-M)_0 = 25.67 \pm 0.13$ mag.

Sextans B:  
Sakai, Madore, \& Freedman (1997) measured the TRGB at
$I = 21.60 \pm 0.10$ mag using ground-based imaging data,
which yields a distance modulus of $(m-M)_0 = 25.61 \pm 0.14$ mag
when adopting $A_I = 0.062$ mag and $M_{I,TRGB} = -4.07$ mag using
the colors of RGB stars tabulated in F00.
More recently, Mendez et al. (2002) have used HST imaging to derive 
a distance of $(m-M)_0 = 25.63 \pm 0.04 \pm 0.18$ mag.
We adopt the average of these measurements,
$(m-M)_0 = 25.63 \pm 0.04$ mag.


NGC~224 (M31):  Mould \& Kristian (1986) used ground-based imaging to
estimate a TRGB magnitude of $I = 20.55 \pm 0.17$ mag for NGC~224.
A more recent study by Durrell, Harris, \& Pritchet (2001) yields the nearly
identical result: $I = 20.52 \pm 0.05$ mag.
We adopt the average of these measurements.
Using $(V-I)_{TRGB} = 1.97 \pm 0.10$ mag and $(V-I)_{-3.5} = 1.7 \pm 0.1$ mag,
as tabulated in F00, we expect $M_{I,TRGB} = -4.07 \pm 0.10$ mag,
producing a distance modulus to NGC~224 of $24.44 \pm 0.11$ mag.


NGC~598 (M33):  Mould \& Kristian, using ground-based data,
derived a TRGB magnitude of $I = 20.95 \pm 0.17$ mag.
Adopting $E(B-V) = 0.04$  mag, and $M_{I,TRGB} = -4.02$ mag
yields a distance modulus of $24.89 \pm 0.20$ mag.
Recently Kim et al.\,(2002) 
used HST/WFPC2 imaging of 10 fields in M33 to derive 
$24.81 \pm 0.04 {^{0.15}_{-0.11}}$ mag.
The average of these measurements yields
$(m-M)_0 = 24.81 \pm 0.04$ mag.


NGC~3109:  
F00 tabulated two sets of data for NGC~3109.
Lee (1993) used deep ground-based $V$ and $I$
imaging to derive $I = 21.55 \pm 0.10$ mag.
Minniti, Zilstra, \& Alonso (1999) derived $I = 21.70 \pm 0.06$ mag also from ground-based data.
Using the colors tabulated in F00, we obtain distance moduli
of $(m-M)_0 = 25.43 \pm 0.14$ mag and $25.60 \pm 0.12$ mag respectively.
More recently, HST observations of M\'endez et al.\,(2002) yielded
$I = 25.52 \pm 0.06$ mag.  Placing 
all of these measurements on the Schlegel et al.\ (1998) reddening
scale gives an average value of $(m - M)_0 = 25.52 \pm 0.05$ mag
for the distance modulus of NGC 3109.

NGC~6822:  Lee et al.\ (1993) derived $(m - M)_0 = 23.46 \pm 0.10$ mag,
Updating the reddening yields a slightly smaller value of 
$23.34 \pm 0.10$ mag.  Subsequently Gallart, Aparicio, \& Vilchez
(1996) derived $23.4 \pm 0.1$ mag, using  reddening values 
estimated from observations of Cepheid variable stars in the same field.
Applying the reddening value from Schlegel et al. (1988) 
yields the same distance modulus $(m-M)_0 = 23.4 \pm 0.1$ mag.
We have adopted the mean of these measurements, $(m-M)_0 = 23.37 \pm 0.07$ mag.

IC~1613:  Freedman (1988) derived $I_{TRGB} = 20.25 \pm 0.15$ mag.
Adopting the colors tabulated in F00,
the corresponding distance modulus is $(m-M)_0 = 24.29 \pm 0.18$ mag.
Subsequently, Cole et al.\ (1999) and Dolphin et al.\ (2001) obtained
HST/WFPC2 imaging of the center and halo of IC~1613, and 
derived TRGB distance moduli of $(m - M)_0 = 24.29 \pm 0.12$ mag and 
24.32 $\pm$ 0.08 mag, respectively.  We adopt the average of these
three values, $(m-M)_0 = 24.31 \pm 0.06$ mag.

WLM:  Two estimates of the TRGB magnitude exist, both based
on ground-based data. Lee et al.\ (1993),
derived $I_{TRGB} = 20.85 \pm 0.10$ mag, while Minniti \& Zijlstra (1997)
found $I_{TRGB} = 20.80 \pm 0.05$ mag.
Averaging these values and applying the RGB colors and reddening from F00
yields a distance modulus of 
$(m-M)_0  = 24.77 \pm 0.09$ mag.


\section{Cepheid Distances}

Published Cepheid distances exist for all galaxies discussed in this
paper, and are listed in Column 2 of Table~3, where the appropriate
references are also given. The assumptions and procedures used in
deriving these distances vary from galaxy to galaxy. For instance, NGC
3031, NGC 3351, NGC 3621, NGC 4258, and the innter field of NGC 5457 were all
observed with HST using the same instrument configuration, WFPC2 and
the F555W ($\sim$ Johnson $V$) and F814W ($\sim$ Johnson $I$) filters.
NGC~5253, IC~4182 and the outer field of NGC~5457 were obtained with
the pre-refurbishment HST/WFC.
Distances to these
galaxies share the same calibration (both zero point and slope) of the
Cepheid PL relation, adopted from Madore \& Freedman (1991, hereafter
MF91). However, the data are not always on a common photometric
system, the latter having been revised several times  since the
installation of WFPC2 on HST. With the exception of the 1988 distance
to IC1613, which preceded MF91, distances from ground-based data
adopt the MF91 PL relation slope, but not always the zero point
(e.g. NGC 3109). Most are based on $BVRI$ data, although for any given
galaxy, not all Cepheids are observed in all four photometric
bands. Only $I-$band data exist for WLM, making it necessary to adopt
an internal reddening to the galaxy (Lee et al. 1993) to transform the
$I-$band distance modulus to a true (de-reddened) one. Finally, in
about half of the cases, the Cepheid sample is truncated at the low
period end prior to calculating a distance, to reduce the effect of
magnitude incompleteness and avoid contamination from overtone
pulsators.

The inhomegeneity in the published Cepheid distances could introduce
artificial trends in our analysis. To create a consistent dataset, we
have converged on the following criteria: all distances must be based
on 1) $VI$ data only, for consistency with the HST sample; 2) for the
HST data, a photometric calibration following Hill et al. (1998); 3) a
common calibration of the PL relation (to be discussed in detail
below); and 4) Cepheids with period between eight  and 100 days, to
exclude overtone pulsators and long period Cepheids, which might define
a different PL relation than their shorter period counterparts. A
longer cut at the short end might be applied to the samples of
Cepheids observed with HST to avoid magnitude incompleteness (see
Ferrarese et al. 2000a and F01).

Because of the differential nature of our comparison, a zero point
shift,  in either the photometric zero point or the Cepheid PL
relation (for instance  due to a change in the LMC distance) has no
effect on the results. The adopted slope of the calibrating Cepheid PL
relation, however, can be critical. The MF91 calibration of the
Cepheids PL relation was based on a sample of 32 LMC Cepheids, mostly
at short periods. A distance modulus of 18.50 mag, a mean and
differential reddening of $E(V-I)=0.13$ and $R=A_V/(A_V - A_I) = 2.45$
(Cardelli, Clayton \& Mathis 1989) were adopted for the LMC. This
calibration has been superseded by the more recent work of the OGLE
consortium (Udalski et al. 1999, hereafter U99), who observed almost
650 Cepheids in the LMC. While the MF91 and U99 calibrations give
identical slopes for the $V-$band PL relation, the slopes in the
$I-$band differ significantly. The impact of this slope change on the
Cepheid distances can be significant -- up to 5\% in some cases -- and
distance dependent (F01). The measured reddening to each Cepheid is
larger when the U99 calibration is used, the more so the longer the
Cepheid's period. Because of observational biases, the period of the
shortest observed Cepheids is generally longer in distant galaxies
than in nearby ones; it follows that, as a general trend, the U99
calibration leads to increasingly larger reddenings, or increasingly
smaller distances, the further away the galaxy under study. Since the
more distant galaxies in our sample happen to be the more metal rich,
it is quite feasible that adopting an incorrect slope for the LMC PL
relation might translate into a spurious metallicity dependence when
Cepheid and TRGB distances are compared.

To assess the impact of systematics, the analysis in the following
sections will be performed  twice, with Cepheid distances derived
using the MF91 and U99 calibration. In all cases, we adopt a distance
modulus of 18.50 mag, and a mean reddening of $E(V-I)=0.13$ mag to the
LMC.  These distances are listed in column 3 and 4 respectively of
Table~3.
For NGC 224, and all of the galaxies observed with HST with
the exception of the two NGC 5457 fields, the distances were adopted
from Table 3 of F01. In the case of distances based on the U99
calibration, F01 adopted the photometric zero points from Stetson et
al. (1998); for consistency, we transformed these distances to the
Hill et al. (1998) photometric system by adding 0.07 mag.
For all other galaxies,
we found it necessary to calculate the distances anew. Additional details
are given in the Appendix.

For some of the galaxies with ground-based distances, Cepheids are observed in more
bands than $V$ and $I$ (for instance $B$ and $R$, see Appendix A). Calculating a 
distance using multi-wavelength data sometimes leads to improvements over fits 
which only use $V$ and $I$ data, especially in the case of  sparsely sampled PL 
relations. Distances using all available photometric bands are 
therefore listed in Table~3, columns 5 and 6; these agree 
identically to the distances tabulated in Columns 3 and 4 when only $V$ and $I$ 
data are available.

\section{Abundances}

Metal abundances for most
extragalactic Cepheids cannot be measured directly. The HST Key Project 
adopted [O/H] nebular abundances derived from spectra of HII regions 
at the same galactocentric distance as the Cepheid fields (Zaritsky, Kennicutt,
\& Huchra 1994, hereafter ZKH; K98; and Ferrarese et al.\ (2000a). Although 
these should provide reasonable estimates of the [Fe/H] stellar abundances 
for relatively massive, luminous, and short lived Cepheids, we note that  
a one-to-one correspondence is not essential: the inferred metallicity 
dependence of the Cepheid PL relation should be valid so long as it is 
applied using a self-consistent nebular abundance scale.

The ZKH "empirical" abundances were derived
from the relative strengths of the [\ion{O}{2}]$\lambda$3726,3729
[\ion{O}{3}]$\lambda$4959,5007, and H$\beta$ emission lines, and
calibrated with a combination of observations and theoretical
nebular photoionization models (e.g., Edmunds \& Pagel 1984;
Kewley \& Dopita 2002).  The adopted abundances are listed in the 2nd
column of Table~4.  
For some metal-poor dwarf galaxies ($Z < 0.3 Z_\odot$), 
direct HII region oxygen abundances derived from electron temperature ($T_e$)
measurements were available and were used instead.

There have been two significant developments in the abundance scale
since the publication of the K98 analysis.  First, improved measurements
of the CNO abundances in the Sun have resulted in a downward revision
of the solar oxygen abundance scale, from $12 + \log~{\rm O/H} = 8.9$
to 8.7 (Allende Prieto et al.\ 2001; Holweger 2001).  All values of
[O/H] in this paper will be referenced to the new lower solar abundance.
Second, recent measurements of $T_e$-based abundances in several galaxies
reveal that the strong-line empirical abundance are systematically higher
than the direct abundances by 0.3--0.5 dex, for HII regions more 
metal-rich than  $Z \sim Z_{LMC}$
(Kennicutt, Bresolin, \& Garnett 2003 and references therein).
This difference affects the formally derived slope of the Cepheid 
metallicity dependence, because adopting direct $T_e$-based abundances
lowers the metal-rich end of the abundance scale without changing 
the abundances adopted for metal-poor HII regions (and Cepheids).
Fortunately this does not significantly change the effects of any 
Cepheid $Z$-dependence on the distance scale, so long as 
the metallicity corrections are applied and calibrated using the same
nebular abundance scale.  The issue is however relevant for understanding
the physical origins of any Cepheid $Z$-dependence, where the absolute
magnitude of the effect is important.  

In order to remain consistent
with K98 and the published extragalactic Cepheid studies, we will continue
to adopt the ZKH abundance scale in our analysis.  However in \S6 we 
also discuss the impact of adopting a $T_e$-based abundance scale on 
the absolute scale of the Cepheid metallicity dependence.

\section{Metallicity Dependence of the Cepheid PL Relation}

The main result of our study is shown in Figure~\ref{figure:metaldep},
which plots the difference between Cepheid and TRGB distance moduli 
as a function of Cepheid [O/H] abundance (for $12 + \log O/H = 8.7$;
the solar abundance is 8.7).
The figure includes two datapoints for M101 (which has
Cepheid measurements in two fields at different metallicities) and one for each of 
the other 16 galaxies in Table~2.

In order to test the sensitivity of our results to the Cepheid samples
and calibration used, Table~5 lists various fits to the different samples.
The first section in Table~5 and Figure~\ref{figure:metaldep}
present the results 
obtained when the TRGB distances (column 2 of Table 2) are compared to the following: 
(1) Cepheid distances as published in the original papers
(column 2 of Table~3); 
(2) Cepheid distances derived using only $V$ and $I$ data, calibrated as in
Madore \& Freedman (1991) (column 3 of Table~3);
(3) Cepheid distances derived using only $V$ and $I$ data, calibrated as in
Udalsky (1999) (column 4 of of Table~3);
(4) Cepheid distances based on multiwavelength fits (when available), calibrated using
MF91(column 5 of Table 3)
(5) Cepheid distances based on multiwavelength fits (when available), calibrated using
Udalski (1999) (column 6 of Table 3). 
In fits (2) and (3) (based on $V$ and $I$ data only),
Sextans~B was excluded, since its distance, measured using only
two Cepheids, is likely unreliable (a conclusion supported by the fact that the
galaxy is a significant outlier in the 2nd and 3rd panel of Figure~\ref{figure:metaldep}).
The least square fits account only for the errors in the distance moduli; 
the uncertainties in the abundances, [O/H], are mostly
systematic in nature and do not affect the fits.


All of the comparisons show a clear trend with metallicity, with Cepheid
residual modulus decreasing with increasing [O/H].  This is in the 
same sense as reported earlier by Freedman \& Madore (1990), Gould (1994),
Sasselov et al.\ (1997), Kochanek (1997), and K98, but now the
dependence is seen much more clearly.  The
slope $\gamma$ and its rms uncertainty, derived from a weighted least-squares fit,
are shown in the upper right corner of each panel.  
We note that the main effect of adopting the Udalski et al.\,(1999) calibration instead of the MF91 scale
is to reduce the average Cepheid distance moduli by about 0.08 mag regardless of the
metallicity of the sample, so that $\gamma$ is unaffected.

Since there is no a priori reason to prefer one dataset of Cepheid                        
distances to another, we take the average of the $\gamma$ values returned                 
by the five fits as our best estimate of the metallicity dependence of the                
Cepheid PL relation. Strictly speaking, the five fits are correlated since                
each one is always applied to the same set of galaxies. However, the                      
methods used are different for the published, the V/I and the                             
multi-wavelength data. In that sense, the estimates are independent                       
because they were derived by applying independent methods. Therefore, we                  
assume the following; 
the two fits to the $V$ and $I$ data alone based on the MF91 and                          
U99 calibrations are correlated, as well as the two fits using the                        
multiple-wavelength data based on the same two calibrations. Thus, we                     
first estimated the average for each of two sets of correlated fits. The                  
uncertainty was chosen to encompass the range of two error bars. Finally,                 
the value of $\gamma$ was derived by taking the weighted average of three                 
independent estimates (from published Cepheid data, averaged V/I data                     
sets,                                                                                     
and averaged multiple-wavelength data sets):                                              

\begin{equation}
\gamma = -0.24 \pm 0.05 {\rm mag~dex^{-1}}
\end{equation}

\noindent 
If we had assumed that all fits are very highly correlated                      
(rather than statistically independent), then the value 
of $\gamma$ would be $-0.23 \pm 0.11$, which agrees well with the value                   
above. Our measurement of $\gamma$ is similar to $\gamma = -0.24                          
\pm 0.16$ derived by K98 using two Cepheid fields in M101, and $\gamma =                  
-0.12 \pm 0.08$ derived from a comparison of Cepheid and TRGB distances                   
for a smaller sample. As reflected in the decreased errorbar, the results                 
presented in this paper are more robust: the K98 analysis suffered from a                 
lack of galaxies with $Z > Z_{LMC}$, and used an indirect TRGB and                        
Cepheid comparison (via different galaxies in the same group) for two of                  
the metal-rich fields. Both shortcomings have been corrected in this                      
study.

The Cepheid and TRGB distances used in our analysis come from a wide
range of ground-based and HST observations, and it is
important to confirm that the trends shown in Figure~\ref{figure:metaldep}
do not arise from biases built into the sample, as due, for instance, to crowding 
effects or photometric scale errors which might affect ground and space based determinations differently.
As discussed in \S\,3, several of the galaxies in our sample have TRGB
distances measured both from the ground and with HST, and the excellent
agreement between most of these measurements offers some assurances as to consistency of the two sets of
measurements.
In the top panel of Figure~\ref{figure:gbvshst} the datapoints
are coded according the source (ground-based or HST) of the Cepheid distances.
Although most low metallicity galaxies were observed from the ground, while most high 
metallicity galaxies were observed with HST, the middle ground is covered by both HST 
and ground-based measurements, and no systematic difference between the two sets is 
evident. Fitting the ground-based data alone yields a metallicity dependence
$\gamma = -0.18 \pm 0.10$, fully consistent with the value of  $-0.24 \pm 0.08$ 
derived for the combined data.

The lower panel of Figure~\ref{figure:gbvshst} shows the same
comparison, but this time with the points coded according to the source of 
the TRGB distances.  Again a consistent trends is seen across the data set.
Excluding the galaxies with only HST TRGB distances yields 
$\gamma = -0.13 \pm 0.12$.
Excluding the galaxies with only ground-based TRGB distances,
$\gamma = -0.23 \pm 0.14$.
This gives us confidence
that our measured metallicity dependence is not an artifact of instrumental effects and/or crowding errors.

Another conceivable source of systematic error in this comparison would
be a residual metallicity dependence in the TRGB distances, which might
masquerade as an effect on the Cepheid distances.  As discussed in 
\S3 there is a weak ($\pm 0.1$ mag) metallicity dependence in the TRGB 
magnitude,
which is calibrated and corrected for using the observed (dereddened)
$V - I$ color of the giants.  The validity of these corrections is
supported by observations of multiple fields in M33 by Kim et al.\,(2002).
For this effect to be significant in our analysis, the halo metal abundances 
would have to correlate systematically with the Cepheid (disk) abundances.
We present such a comparison in Figure~\ref{figure:zcompare}, where
we plot the dereddened $V - I$ color of the red giants against the 
adopted [O/H] abundance of the Cepheids; for galaxies with multiple
TRGB measurements, we show each data point separately. 
This comparison, unfortunately, shows a slight correlation between
the two sets of abundances; the galaxies with redder RGB stars tend to
be more metal-rich in the oxygen abundance scale as well.
A least-squares fit to the data yields a slope of $0.16 \pm 0.07$,
which is not consistent with being zero.
In order to estimate whether this has amplied the metallicity effect
we have observed, the least-squares fitting is carried out by excluding
three galaxies whose RGB colors exceed $(V-I)>1.8$ mag. 
For all five samples (equivalent to top five rows in Table~5),
the slopes are consistent with those estimated using the
whole sample.  We obtain, for e.g., $\gamma = 0.23 \pm 0.07$ and $-0.25 \pm 0.08$
for the MF91 and U99 multi-wavelength samples, respectively.
Using all galaxies, we had obtained $\gamma = -0.23 \pm 0.08$ and
$-0.24 \pm 0.08$.  
Thus, the slight correlation between the Pop~I and Pop~II abundances
in the galaxies in our sample does not appear to be the cause of the
metallicity dependence of Cepheid variable stars.



Another important question about the metallicity dependence
is whether it is present across the entire range of Cepheid abundances,
or is only important in metal-rich objects.  For example, if the effect
on derived distance moduli were a linear function of metal fraction, 
as was parametrized by Choisi et al.\,(1993), it would have a negligible
influence for dwarf galaxies with $Z \ll Z_{LMC}$, but could be  very
important effect in luminous metal-rich galaxies.  Our data are not
of sufficient quantity or quality to constrain unambigously the 
functional form of the $Z$-dependence.  However, we can examine the 
plausibility of a continuous metallicity dependence by fitting 
the data points for $Z \le Z_{LMC}$ and $Z \ge Z_{LMC}$ separately.
The resulting dependences are nearly identical:
$\gamma = -0.17 \pm 0.13$ for $Z \le Z_{LMC}$ (12 points) and 
$-0.22 \pm 0.19$ for $Z \ge Z_{LMC}$ (8 points).  Our data
are consistent with a continuous logarithmic metallicity dependence,
but the result has marginal statistical significance.

Finally, we note that distances of two galaxies, Sextans~A and 
NGC~5457 (inner field),
vary significantly among the five estimates listed in Table~3.
For example, the distances of the inner field of NGC~5457 
vary from $28.93 \pm 0.11$ up to $29.21 \pm 0.09$ mag.
In order to assess how sensitive the value of $\gamma$ is to
some specific galaxy distances, 
we have estimated $\gamma$ for samples 
including and excluding these galaxies.
We find  that $\gamma$ is stable; its value, when estimated
excluding Sextans~A and N5457 (inner), agrees within
1$\sigma$ of the result quoted above.

\section{Discussion}

The results of this analysis provide the strongest evidence to date
for a non-negligible dependence of Cepheid distances on metal abundance.
Our best estimate of the magnitude of this dependence is 
$\gamma = -0.24 \pm 0.05$ mag~dex$^{-1}$,
when referenced to the Zaritsky et al.\,(1994) HII region metallicity
scale (we consider the effects of adopting a different metal abundance
scale below).  This result is consistent with the
dependence measured from a direct comparison of  metal-rich and 
metal-poor Cepheid fields in M101 ($\gamma = -0.24 \pm 0.16$;
K98).
In the remainder of this section, 
we explore the consequences of such a $Z$-dependence on
the calibration of the distance scale as a whole and H$_0$.  

The consequences of a Cepheid metallicity dependence of roughly 
this magnitude on the calibration of several extragalactic standard
candles was explored in detail in the final series of papers from the 
HST H$_0$ Key Project (Sakai et al.\,2000, 
Ferrarese et al.\,2000b; Gibson et al.\,2000;
Kelson et al.\,2000; Mould et al.\,2000; F01).
We have summarized these results in Table~6, which shows the net effect
of a Cepheid metallicity dependence of 0.20 mag~dex$^{-1}$ on the zeropoint 
calibrations of the secondary distance indicators used by the Key Project 
team.  These are expressed in terms of the luminosity zeropoints and
on the mean net change in the derived distances for the Key Project samples.

A Cepheid $Z$-dependence in 
the direction measured here causes {\it all} of the secondary distance
scales to be systematically underestimated (thus leading to an over-estimate
of H$_0$).  This is because the PL relation is calibrated with a relatively
metal-poor galaxy, the LMC.  The magnitude of the effect is slightly 
different for the different secondary distance indicators, but for 
$\gamma = -0.20$ mag~dex$^{-1}$ it is significant but small, lowering
the net value of H$_0$ by 3.5\%, or about 2.5 km\,s$^{-1}$\,Mpc$^{-1}$
for H$_0$ = 72 km\,s$^{-1}$\,Mpc$^{-1}$ (F01).
This correction already has been incorporated into the value given above.

As mentioned in \S\,6 the absolute slope of the Cepheid $Z$-dependence is also
sensitive to the metallicity scale adopted.  As an illustration of this
point Figure~\ref{figure:znew} shows the same Cepheid vs TRGB comparison
as Figure~\ref{figure:metaldep}, but with the metal abundances adjusted
to agree with the electron temperature based HII region abundances in
Kennicutt et al.\,(2003).  As discussed earlier this has the effect
of preferentially reducing the metallicities of the most metal-rich
Cepheid fields, and the result is a somewhat ($\sim$25\%) steeper 
$Z$-dependences, with an average $\gamma = -0.31 \pm 0.09$ mag~dex$^{-1}$.
Note however that adopting this different abundance scale {\it would have
an identical effect on the distance scale} as given in Table~6, 
because the effect of the steeper $Z$-dependence would be canceled
by a correspondingly narrower abundance range in the calibrating galaxies;
in other words as long as the metallicity corrections are applied
using a consistent abundance scale, the precise calibration of the 
metallicity scale is not important.  Of course the absolute slope
of the dependece is important for understanding the physical origins
of the period-luminosity dependence of the Cepheid variable stars.

In \S3, it was suggested that because the study presented in this
paper is a {\it differential} test, it would not matter which
TRGB calibration is used.  We test this assumption by examining the
metallicity dependence using two independent calibrations.
The first one is that by Lee et al. (1993) which was used throughout
this paper. The second calibration is that by Salaris \& Cassisi (1998),
which is based on the stellar evolution models.  The dominant
different between the two calibration is that the theoretical model
by Salari \& Cassisi predicts the TRGB magnitude $\sim 0.1$ mag brighter
than the empirical calibration of Lee et al. 
The authors suggest that the difference arises from the fact that
the globular cluster samples used in the empirical calibration may be
missing the brightest RGB stars due to the small number statistics, 
and thus systematically dimming the TRGB magnitude.
In Figure~\ref{figure:zpcomp},  we show two correlations, one using
the Lee et al. calibration, and the other based on Salaris \& Cassisi 1998.
As expected, the zero points of the two correlations differ by $\sim$ 0.1 mag.
For the MF91, multi-wavelength sample, using the theoretical calibration,
we obtain $\gamma = -0.26 \pm 0.08$, which agrees well with the fit using
the empirical, Lee et al. calibration, $\gamma = -0.23 \pm 0.08$.
In summary, we emphasize again that the results shown in this paper
are based on {\it differential} tests, and as indicated by our simple 
comparison, the value of $\gamma$ should not be affected by the use of
another TRGB calibration.

Finally, our measurment of the metallicity dependence
cannot distinguish between a variation in the zero point
or in the slope.
As discussed in Section~4, the slope of the Cepheid PL relation is
not always well determined; the MF91 and U99 calibrations in fact 
yield I-band slopes that are significantly different from each other.
Thus, there is a need to check if the slope is the cause of 
the metallicity dependence of the Cepheid variables.
A detailed study is beyond the scope of this paper;
here a simple exercise is carried out to examine the effect of the slope,
by calculating the mean period of Cepheid variable sample for each galaxy.

When the mean periods are plotted against the metallicities,
we find that there are two groupings: one around $12+log(O/H) \sim 8.7$
and mean $P \sim 1.4$, and the other one at $12+ \log (O/H) \sim 7.7$ and
mean $P \sim 1.1$.  That is, one group at high Z corresponds to the longer mean
period, and the second one at low Z corresponds to the shorter mean period.
The second low-Z, shorter period group consists of four galaxies.
The mean Cepheid periods were also calculated for all the galaxies
used as the calibrators for the Tully-Fisher relation.  
This is especially
important to check if slope slope is responsible for the metallicity
dependence of the Cepheids, affecting the calibration of the secondary
distance indicators and finally the value of H$_0$. 
The Tully-Fisher calibrators all lie in the
high-Z, long mean period group.
If the metallicity affects the slope of the Cepheid PL relation,
then we might need to exclude those galaxies whose mean period
is significantly different from others. 
Thus, excluding four galaxies that have low mean periods,
the metallicity dependence, $\gamma$, was re-calculated.
For the multiple-wavelength, MF91 calibration sample, 
$\gamma = -0.25 \pm 0.09$,
which agrees well with the value estimated using all galaxies ($\gamma=-0.24$).
Therefore, to first order, the slope of the Cepheid PL relation
does not appear to affect the metallicity dependence.

\acknowledgements

This project was made possible by an allocation of observing time
on HST (program GO-8584).  We gratefully acknowledge the assistance
of Ray Lucas in carrying out this program, and the financial support
of grant HST-GO-08584.  This research has made use of the NASA's
Astrophysics Data System.
We would also like to thank to anonymous referee for suggestions
which helped in improving this paper.

\section{Appendix A: Comments on galaxies for which new Cepheid distances
have been calculated in this paper.}

Sextans A: a total of 10 Cepheids,  six with period longer
than 8 days, is known in this galaxy (Sandage \& Carlson 1984; Piotto,
Capaccioli \& Pellegrini 1994). Sakai, Madore \& Freedman (1996)
recalibrated the Cepheid photometry by comparison with new CCD $BVRI$
data for non-variable stars in the field, and calculated the distance
reported in column 2 of Table 4. 

Sextans B: Sandage \& Carlson (1984) discovered seven Cepheids in this
galaxy. Three of these were confirmed by Piotto, Capaccioli \&
Pellegrini (1994) based on ground based $BVRI$ data. The same authors
discovered four shorter period Cepheids, and used the entire sample to
calculate a distance modulus of $25.63 \pm 0.21$ mag. Unfortunately,
$VI$ magnitudes are measured for only three Cepheids, one 
with period shorter than eight days. The distances listed in columns 3 and 4 are therefore based on only two Cepheids, for which we
adopt the data from Piotto, Capaccioli \& Pellegrini (1994).

NGC 300: Eighteen Cepheids were discovered by Graham (1984) based on
photographic data; CCD $BVRI$ photometry was obtained for 16 of these
by Freedman et al. (1992) and used to calculate the distance reported
in Table 4 (col. 2). Recently, Pietrzy\'nski et al. (2002) recovered
and refined the magnitudes and periods for all of Graham's variable
stars, using $BV$ ground based CCD data. This study revealed that
three of the Cepheids used by Freedman et al. (1992) were
blended. These were excluded for the purpose of this paper; the distance
listed in columns 3 and 4 of Table 4
was calculated from the remaining 13 Cepheids using the periods and
$V-$band magnitudes from Pietrzy\'nski et al. (2002), and the $I-$band
magnitudes from Freedman et al. (1992).

NGC 598: Column 2 of Table 4 lists the distance published by Freedman,
Wilson \& Madore based on $BVRI$ CCD photometry of 19 Cepheids
originally discovered by Hubble (1926). The distance moduli calculated
in this paper are based on all Cepheids given a quality index of {it a, b}
or {\it c} in Freedman, Wilson \& Madore (1991), for which both $V$ and $I$
magnitudes are available. 

NGC 3109: $B$ and $V$ photometry was obtained for eight Cepheids by
Capaccioli et al. (1992). Subsequently, Musella, Piotto \& Capaccioli
(1997) extended the photometry to include $R$ and $I$ data,
and discovered 16 additional Cepheids, calculating a distance modulus
of $25.67 \pm 0.16$ mag. Seven of these Cepheids have period longer
than eight days and $VI$ photometry, and were used in calculating the
distances listed in columns 3 and 4 of Table 4. 

NGC 5457 inner: an inner field in M101 was studied by Stetson et
al. (1998) as part of the HST Key Project on the Extragalactic
Distance Scale. In this paper, a distance is calculated using a total
of 61 bona-fide, high quality Cepheids for which Table 4 of Stetson et
al. (1998) lists a quality index larger than 2 under either `Image
Quality' or `Light Curve Quality'

NGC 5457 outer: HST data were obtained as part of the HST Key Project on
the Extragalactic Distance Scale (Kelson et al 1996). The distance
calculated in this paper makes  use of all 29 Cepheids from the original
study.

NGC 6822: Based on photographic data, Kayser (1967) identified 13
Cepheids; CCD data, unfortunately only in the $r$-band, exist for six
of these (Schmidt \& Spear 1989). Photometric transformation from
Kayser's magnitudes to the $BVRI$ system were computed by  Gallart et
al. (1996) and applied to eight of Kayser's Cepheids. Six of these,
with well determined periods, were then used to calculate the distance
reported in column 2 of Table 4. We used the same sample of six
Cepheids to calculate the distance to NGC 6822 listed in columns 3 and 4 of
Table 4. 

IC 1613: The first discovery of Cepheids in this galaxy dates back to
the work of Baade and Hubble, later published by Sandage (1971) and
Carlson \& Sandage (1990). New $BVRI$ CCD data were published for 11 of
the original 24 Cepheids by Freedman (1988) and used to derive the
distance modulus listed in column 2 of Table 4. For the purpose of
this paper, we have retained the four Cepheids with period between
eight and 100 days; periods and $VI$ magnitudes are from Freedman
(1988). 

WLM: A distance to this galaxy was calculated by Lee et al. (1993)
based on $I-$band data, adopting an absorption  $A(I) = 0.04$
mag. Furthermore, all of the Cepheids have periods less than eight
days, preventing us from calculating a new distance for this galaxy.

{}


\begin{deluxetable}{lcccc}
\tabletypesize{\scriptsize}
\label{table:observations}
\tablecaption{HST/WFPC2 Observations}
\tablewidth{0pc}
\tablecolumns{5}
\tablehead{
\colhead{Galaxy} &
\colhead{RA (h:m:s)} &
\colhead{DEC (\dg:\mn:\scn)} &
\colhead{Filter} &
\colhead{Exposure} \cr
\colhead{} &
\colhead{(J2000)} &
\colhead{(J2000)} &
\colhead{} &
\colhead{(sec)} \cr
}
\startdata 
IC~4182  &  13:05:49.5 & 37:36:18  & F555W & 1300 x 2 \cr
         &             &           & F814W & 1300 x 2 \cr
NGC~300  &  00:54:53.5 & -37:41:00 & F555W & 500 x 2 \cr
         &             &           & F814W & 500 x 2 \cr
NGC~3031 & 09:55:33.2  & 69:03:55  & F555W & 1100 x 2 \cr
         &             &           & F814W & 1100 x 2 \cr
NGC~3351 &  10:43:57.8 & 11:42:14  & F555W & 1200 x 4, 1300 x 8 \cr
         &             &           & F814W & 1200 x 6, 1300 x 14 \cr
NGC~3621 & 11:18:16.0  & -32:48:42 & F555W & 1300 x 4 \cr
         &             &           & F814W & 1300 x 6 \cr
NGC~5457 & 14:03:12.5 & 54:20:55   & F555W & 1300 x 2, 1400 x 2 \cr
         &            &            & F814W & 1300 x 2, 1400 x 4 \cr
NGC~5253\tablenotemark{a}& 13:39:55.9 & -31:38:24  & F547M & 200 x 2, 600 x 2 \cr
         &            &            & F814W & 180 x 2, 400 x 2 \cr
\enddata
\tablenotetext{a}{HST Archival data from program ID 6524 (PI: Calzetti)}.

\end{deluxetable}

\begin{deluxetable}{lccl}
\tablecaption{Summary of Adopted TRGB Distances}
\tabletypesize{\scriptsize}
\tablewidth{0pc}
\tablecolumns{4}
\tablehead{
\colhead{Galaxy} &
\colhead{TRGB Modulus\tablenotemark{a}} &
\colhead{$A_I$\tablenotemark{b}} &
\colhead{References} \cr 
}

\startdata
LMC	         & 18.59 $\pm$ 0.09 &   0.15    & 1 \\
SMC              & 18.99 $\pm$ 0.08 &   0.27    & 2 \\
Sextans~A        & 25.67 $\pm$ 0.13 &   0.09    & 3,4 \\
Sextans~B        & 25.63 $\pm$ 0.04 &   0.06    & 5,6 \\
NGC~224 (M31)    & 24.47 $\pm$ 0.11 &   0.15    & 7,9 \\
NGC~300          & 26.65 $\pm$ 0.07 &   0.03    & 10  \\
NGC~598 (M33)    & 24.81 $\pm$ 0.04 &   0.08    & 7,11 \\
NGC~3031 (M81)   & 28.03 $\pm$ 0.12 &   0.16    & 10    \\
NGC~3109         & 25.52 $\pm$ 0.05 &   0.13    & 6,12,13     \\
NGC~3351         & 30.39 $\pm$ 0.13 &   0.05    & 10 \\
NGC~3621         & 29.36 $\pm$ 0.11 &   0.16    & 10 \\
NGC~5253         & 27.88 $\pm$ 0.11 &   0.11    & 10 \\
NGC~5457 (M101)  & 29.42 $\pm$ 0.11 &   0.02    & 10 \\
NGC~6822         & 23.37 $\pm$ 0.07 &   0.46    & 7,14 \\
IC~1613          & 24.31 $\pm$ 0.06 &   0.05    & 7,15,16 \\
IC~4182          & 28.25 $\pm$ 0.06 &   0.03    & 10      \\
WLM              & 24.77 $\pm$ 0.09 &   0.07    & 7,8 \\

\tablenotetext{a} {True distance modulus $(m-M)_0$, including extinction correction.  
The quoted uncertainties include observational errors but exclude systematic 
errors in the zeropoint of the TRGB scale.}

\tablenotetext{b} {Adopted $I$-band extinction in mag, from Schlegel et al. (1998)
except for the LMC and NGC 6822 (\S 3.4).}

\tablerefs{
(1) Sakai et al.\,2000; (2) Cioni et al.\,2000; 
(3) Sakai et al.\,1996; (4) Dolphin et al.\,2003; (5) Sakai et al.\,1997; 
(6) M\'endez et al.,2002; (7) Lee et al.\,1993; (8) Minniti \& Zijlstra 1997;
(9) Durrell et al.\,2001; (10) this paper; (11) Kim et al.\,2002;
(12) Lee 1993; (13) Minniti et al.\,1999; (14) Gallart et al.\,1996;
(15) Cole et al.\,1999; (16) Dolphin et al.\,200}

\enddata
\end{deluxetable}

\begin{deluxetable}{lcccccc}
\label{table:cepheiddistances}
\tablecaption{Cepheid Distances}
\tabletypesize{\scriptsize}
\tablewidth{0pc}
\tablehead{
\colhead{Galaxy Name} &
\colhead{$\mu_{pub}$ (mag)\tablenotemark{a}} &
\colhead{$\mu_{MF91}$ (mag)\tablenotemark{b}} &
\colhead{$\mu_{U99}$ (mag)\tablenotemark{c}} &
\colhead{$\mu_{multi-\lambda,MF91}$ (mag)\tablenotemark{d}} &
\colhead{$\mu_{multi-\lambda,U99}$ (mag)\tablenotemark{e}} &
\colhead{References\tablenotemark{f}}\cr
}
\startdata
LMC         & 18.50 $\pm$ 0.10 & 18.50 $\pm$ 0.10 & 18.50 $\pm$ 0.10 & 18.50 $\pm$ 0.10 & 18.50 $\pm$ 0.10 &  1, 1, 1 \cr
SMC         & 18.99 $\pm$ 0.05 & \nodata          & \nodata          & 18.99 $\pm$ 0.05 & 18.99 $\pm$ 0.05 &  2, -, - \cr
Sextans A   & 25.85 $\pm$ 0.15 & 25.62 $\pm$ 0.09 & 25.66 $\pm$ 0.14 & 25.85 $\pm$ 0.15 & 25.79 $\pm$ 0.15 &  3, 4, 4 \cr
Sextans B   & 25.63 $\pm$ 0.21 & 26.43 $\pm$ 0.15 & 26.37 $\pm$ 0.09 & 25.63 $\pm$ 0.21 & 25.63 $\pm$ 0.21 &  5, 4, 4 \cr
NGC~224     & 24.42 $\pm$ 0.12 & 24.41 $\pm$ 0.08 & 24.38 $\pm$ 0.05 & 24.41 $\pm$ 0.08 & 24.38 $\pm$ 0.05 &  6, 1, 1 \cr
NGC~300     & 26.70 $\pm$ 0.11 & 26.63 $\pm$ 0.06 & 26.53 $\pm$ 0.05 & 26.63 $\pm$ 0.06 & 26.53 $\pm$ 0.05 &  7, 4, 4 \cr
NGC~598     & 24.64 $\pm$ 0.09 & 24.56 $\pm$ 0.11 & 24.47 $\pm$ 0.11 & 24.56 $\pm$ 0.11 & 24.47 $\pm$ 0.11 &  8, 4, 4 \cr
NGC~3031    & 27.80 $\pm$ 0.20 & 27.75 $\pm$ 0.07 & 27.75 $\pm$ 0.08 & 27.75 $\pm$ 0.07 & 27.75 $\pm$ 0.08 &  9, 1, 1 \cr
NGC~3109    & 25.67 $\pm$ 0.16 & 25.56 $\pm$ 0.27 & 25.54 $\pm$ 0.28 & 25.56 $\pm$ 0.27 & 25.54 $\pm$ 0.28 &  10, 4, 4 \cr 
NGC~3351    & 30.01 $\pm$ 0.19 & 30.03 $\pm$ 0.10 & 29.92 $\pm$ 0.09 & 30.03 $\pm$ 0.10 & 29.92 $\pm$ 0.09 &  11, 1, 1 \cr
NGC~3621    & 29.13 $\pm$ 0.18 & 29.21 $\pm$ 0.06 & 29.15 $\pm$ 0.06 & 29.21 $\pm$ 0.06 & 29.15 $\pm$ 0.06 &  12, 1, 1 \cr
NGC~5253    & 28.06 $\pm$ 0.30 & 27.60 $\pm$ 0.10 & 27.63 $\pm$ 0.14 & 27.60 $\pm$ 0.10 & 27.63 $\pm$ 0.14 &  13, 1, 1 \cr
NGC~5457in  & 29.05 $\pm$ 0.14 & 29.01 $\pm$ 0.11 & 28.93 $\pm$ 0.11 & 29.21 $\pm$ 0.09 & 29.14 $\pm$ 0.09 &  14, 4, 4 \cr
NGC~5457out & 29.34 $\pm$ 0.17 & 29.34 $\pm$ 0.08 & 29.24 $\pm$ 0.08 & 29.34 $\pm$ 0.08 & 29.24 $\pm$ 0.08 &  15, 4, 4 \cr
NGC~6822    & 23.49 $\pm$ 0.08 & 23.40 $\pm$ 0.07 & 23.30 $\pm$ 0.07 & 23.49 $\pm$ 0.08 & 23.39 $\pm$ 0.08 &  16, 4, 4 \cr
IC~1613     & 24.29 $\pm$ 0.14 & 24.21 $\pm$ 0.31 & 24.17 $\pm$ 0.33 & 24.29 $\pm$ 0.14 & 24.25 $\pm$ 0.14 &  17, 4, 4 \cr
IC~4182     & 28.36 $\pm$ 0.09 & 28.36 $\pm$ 0.06 & 28.35 $\pm$ 0.06 & 28.36 $\pm$ 0.06 & 28.35 $\pm$ 0.06 &  13, 4, 4 \cr
WLM        & 24.92 $\pm$ 0.21 & \nodata	     & \nodata	             & 24.92 $\pm$ 0.21 & 24.92 $\pm$ 0.21 &  18, -, - \cr
\enddata

\tablenotetext{a}{Original published distance (see references below)}
\tablenotetext{b}{Distance based on the Madore \& Freedman (1991) calibration of the
Cepheid PL relation, applied to $VI$  data only.}
\tablenotetext{c}{Distance based the Udalski et al. (1999) calibration of the 
Cepheids  PL relation, applied to $VI$ data only.}
\tablenotetext{d}{Distance based the Madore \& Freedman (1991) calibration of the 
Cepheids  PL relation, using data in all available photometric bands (see \S 4).}
\tablenotetext{e}{Distance based the Udalski et al. (1999) calibration of the 
Cepheids  PL relation, using data in all available photometric bands (see \S 4).}
\tablenotetext{f}{References for Columns 2, 3, and 4}

\tablerefs{(1) Freedman et al. 2001;
(2) Welch et al. 1987;
(3) Sakai, Madore \& Freedman 1996; 
(4) this paper; 
(5) Piotto, Capaccioli \& Pellegrini 1994; 
(6) Freedman \& Madore 1990; 
(7) Freedman et al. 1992; 
(8) Freedman, Wilson \& Madore 1991; 
(9) Freedman et al. 1994; 
(10) Musella et al. 1997; 
(11) Graham et al. 1997;
(12) Rawson et al. 1997;
(13) Saha et al. 1994; 
(14) Stetson et al. 1998; 
(15) Kelson et al. 1996; 
(16) Gallart et al. 1996;  
(17) Freedman 1988; 
(18) Lee et al. 1993.}

\end{deluxetable}

\begin{deluxetable}{lcc}
\label{table:abundances}
\tablecaption{Cepheid Abundance}
\tablewidth{0pc}
\tablehead{
\colhead{Galaxy} &
\colhead{$12 + \log \mbox{(O/H)}$} &
\colhead{$12 + \log \mbox{(O/H)}$}  \cr
\colhead{Metallicity} &
\colhead{ZKH} &
\colhead{T$_e$ based}  \cr}
\startdata
LMC        &  8.50    &  8.34   \cr
SMC        &  7.98    &  7.98   \cr
Sextans A  &  7.49    &  7.49   \cr
Sextans B  &  7.56    &  7.56   \cr
NGC224     &  8.98    &  8.68   \cr
NGC300     &  8.35    &  8.35   \cr
NGC598     &  8.82    &  8.55   \cr
NGC3031    &  8.75    &  8.5    \cr
NGC3109    &  8.06    &  8.06   \cr
NGC3351    &  9.24    &  8.85   \cr
NGC3621    &  8.75    &  8.5    \cr
NGC5253    &  8.15    &  8.15   \cr
NGC5457in  &  9.20    &  8.7    \cr
NGC5457out &  8.50    &  8.23   \cr
NGC6822    &  8.14    &  8.14   \cr
IC1613     &  7.86    &  7.86   \cr
IC4182     &  8.40    &  8.20   \cr
WLM        &  7.74    &  7.74   \cr

\enddata
\end{deluxetable}

\begin{deluxetable}{lccl}
\label{table:metallicity}
\tablecaption{Metallicity Dependence\tablenotemark{a}}
\tabletypesize{\scriptsize}
\tablewidth{0pc}
\tablehead{
\colhead{Sample} &
\colhead{Zero point} &
\colhead{Slope $\gamma$ }  &
\colhead{Notes} \cr}

\startdata

Published Cepheid distances            &  2.09 $\pm$ 0.67  &  -0.25 $\pm$ 0.08 &  \cr
MF91 calibration; $V,I$ data           &  1.61 $\pm$ 0.73  &  -0.20 $\pm$ 0.09 &  \cr
U99 calibration; $V,I$ data            &  1.92 $\pm$ 0.76  &  -0.25 $\pm$ 0.09 &  \cr
MF91 calibration; multi-$\lambda$ data &  1.81 $\pm$ 0.63  &  -0.23 $\pm$ 0.08 &  \cr
U99 calibration; multi-$\lambda$ data  &  1.87 $\pm$ 0.69  &  -0.24 $\pm$ 0.08 &  \cr

  &  &  \cr
Ground Cepheids     & 1.42 $\pm$ 0.82  &  -0.18 $\pm$ 0.10  & Cepheid data based on ground-based observations  \cr
HST Cepheids        & 1.08 $\pm$ 1.35  &  -0.15 $\pm$ 0.15  & Cepheid data based on HST observations  \cr

  &  &  \cr

Ground TRGB         & 1.08 $\pm$ 0.99  &  -0.13 $\pm$ 0.12 & TRGB data based on ground-based observations \cr
HST TRGB            & 1.82 $\pm$ 1.17  &  -0.23 $\pm$ 0.14 & TRGB data based on HST observations \cr

  &  &  \cr

Revised Metal Scale & 2.49 $\pm$ 0.71  &  -0.31 $\pm$ 0.09 & Using the $T_e$ metallicity scale \cr

  &  &  \cr
Low Z  ($<$8.50)    &  1.35 $\pm$ 1.07  &  -0.17 $\pm$ 0.13 &  \cr
High Z ($>$8.50)    &  1.78 $\pm$ 1.67  &  -0.22 $\pm$ 0.19 &  \cr

\enddata
\tablenotetext{a}{$(m-M)_{\mbox{\tiny Ceph}} - (m-M)_{\mbox{\tiny TRGB}}$ = Zero Point + Slope $\times$ O/H.}
\end{deluxetable}


\begin{deluxetable}{lrrc}
\label{table:effects}
\tablecaption{Effects on Distance Indicators and H$_0$}
\tablewidth{0pc}
\tablecolumns{4}
\tablehead{
\colhead{Method} &
\colhead{$\delta$M\tablenotemark{a}} &
\colhead{$\delta$d/d\tablenotemark{b}} & 
\colhead{References} \cr
}
\startdata
Tully-Fisher Relation & $-$0.062 mag & +0.029 & 1,5  \cr 
Surface Brightness Fluctuations & $-$0.089 mag & +0.042 & 2,5  \cr
SN Ia Peak Brightness & $-$0.078 mag & +0.037 &
 3,5  \cr
Fundamental Plane & $-$0.106 mag & +0.050 &
   4,5  \cr
Key Project Combined & $-$0.075 mag & +0.035 &
   5  \cr
\enddata
\tablenotetext{a}{Net effect of Cepheid metallicity dependence $\gamma$ =
-0.20 mag~dex$^{-1}$ on absolute magnitude zeropoint calibration.}
\tablenotetext{b}{Net effect of Cepheid metallicity dependence $\gamma$ =
-0.20 mag~dex$^{-1}$ on distance scale zeropoint, expressed as a fraction.}
\tablerefs{(1) Sakai et al.\,(2000); (2) Ferrarese et al.\,(2000);
(3) Gibson et al.\,(2000); (4) Kelson et al.\,(2000); 
(5) Freedman et al.\,(2001)}
\end{deluxetable}


\figcaption[f1.ps]
{Footprints showing the placement of the WFPC2 on each galaxy.
{\bf See accompanying jpg file: 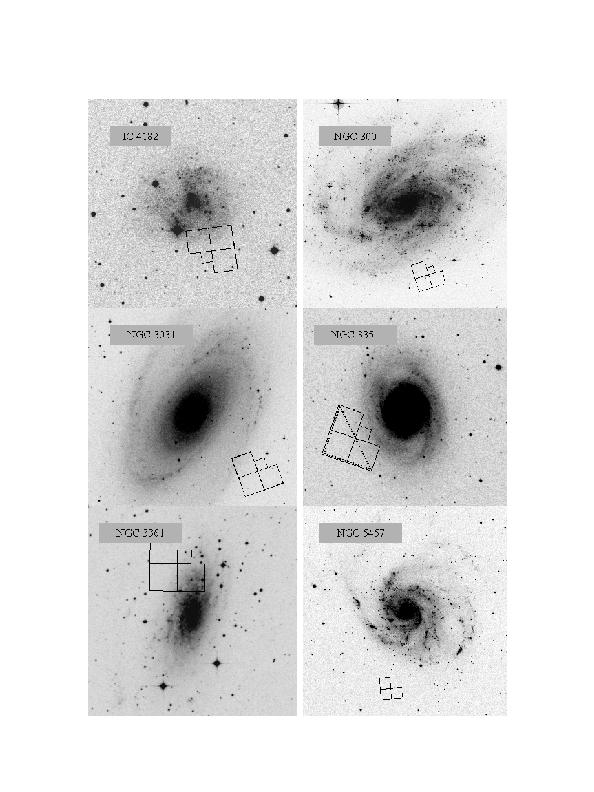}
\label{figure:footprints}}

\figcaption[f2.ps]
{The template luminosity function used in our cross-correlation method
for estimating the TRGB distance.
{\bf See accompanying jpg file: 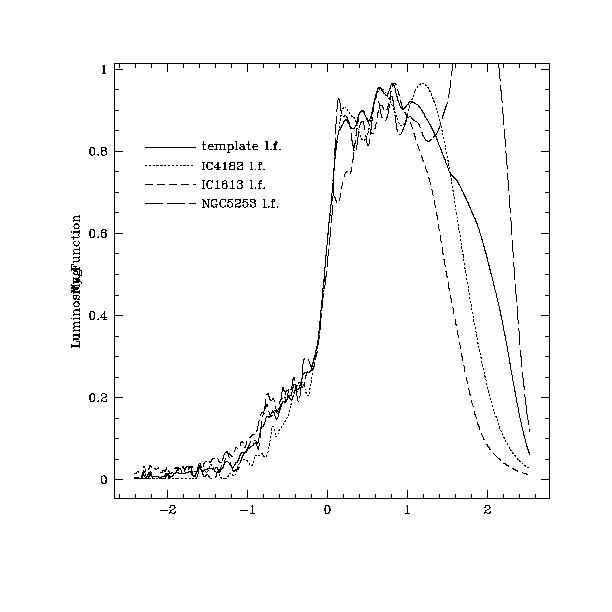}
\label{figure:templatelf}}

\figcaption[f3.ps]
{IC4182: An $I$ vs $(V-I)$ CMD of all stars found in the WFPC2 observations
of IC~4182 {\it (top left)}, and those found on WF2 and 3 chips {\it (bottom left)}.
The linear {\it (top right)} and logarithmic luminosity functions and corresponding
edge-detection filter response functions are also shown.
{\bf See accompanying jpg file: 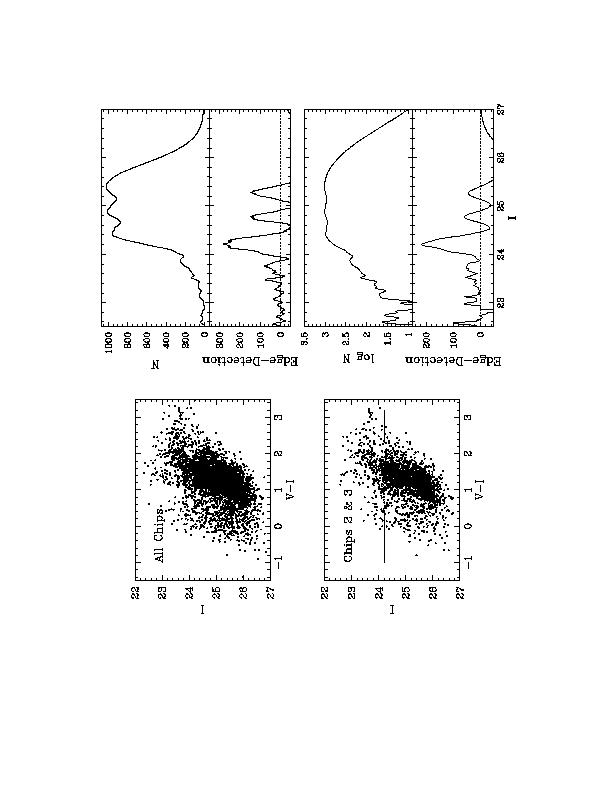}
\label{figure:ic4182tip}}

\figcaption[f4.ps]
{NGC5253 : The WFPC2 image of NGC~5253. 
The halo stars located outside the ellipse were used to determine the
distance to NGC~5253.
{\bf See accompanying jpg file: 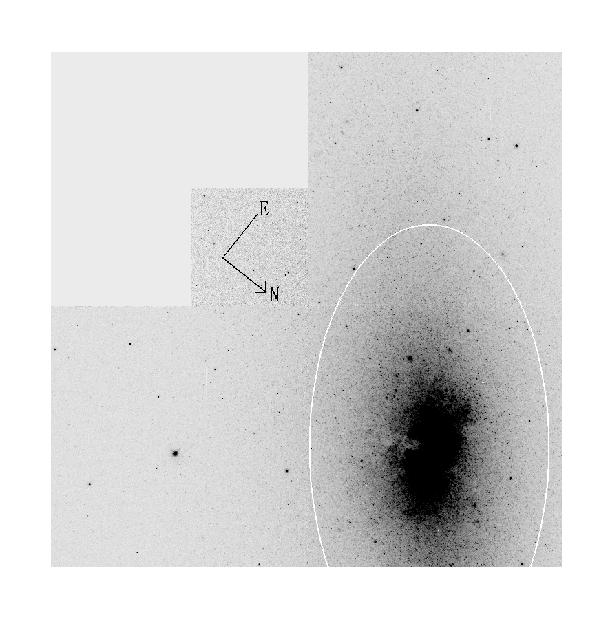}
\label{figure:n5253wfpc}}

\figcaption[f5.ps]
{NGC5253 : An $I$ vs $(V-I)$ CMD of all stars found in the WFPC2 observations
of NGC~5253 {\it (top left)}, and those found in the halo,  outside the
outer ellipse {\it (bottom left)}.
The linear and logarithmic luminosity functions and corresponding
edge-detection filter response functions are also shown {\it (right)}.
{\bf See accompanying jpg file: 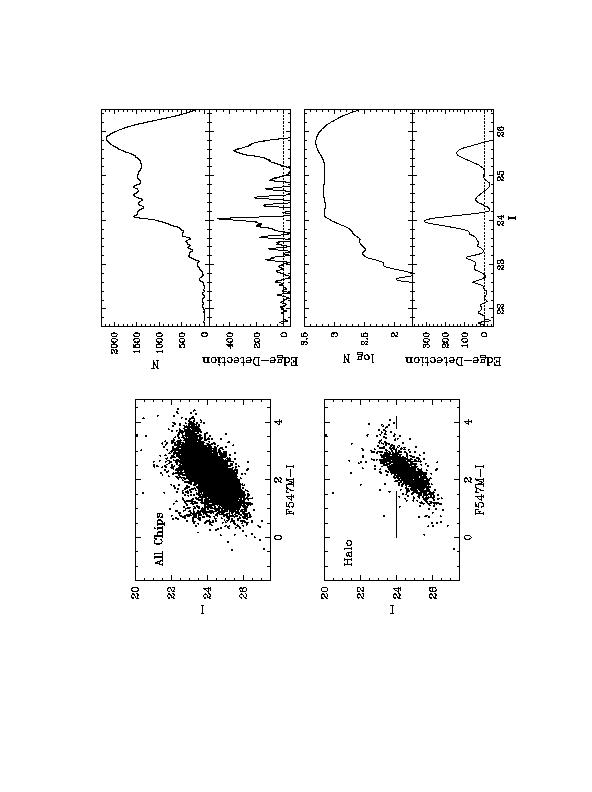}
\label{figure:n5253tip}}

\figcaption[f6.ps]
{NGC~300 : {\it (left): } An $I$ vs $(V-I)$ CMD of all stars found in the WFPC2 observations
of NGC~4182.
{\it (right):} The linear and logarithmic luminosity functions and corresponding
edge-detection filter response functions are also shown. 
{\bf See accompanying jpg file: 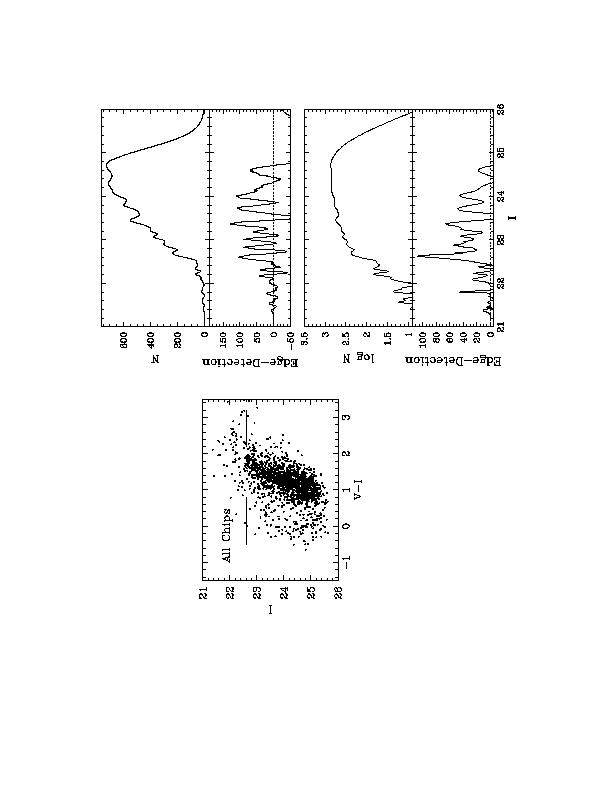}
\label{figure:n300tip}}

\figcaption[f7.ps]
{NGC~3031: An $I$ vs $(V-I)$ CMD of all stars found in the WFPC2 observations
of NGC~3031 {\it (top left)}, and those found on WF3 and 4 chips {\it (bottom left)}.
The linear and logarithmic luminosity functions and corresponding
edge-detection filter response functions are also shown {\it (right top and middle)}. 
Also shown here is the logarithmic luminosity function and its corresponding
edge-detection filter response function for the sample consisting of star
detected on the I-band image, but not necessarily on the V-band image.
{\bf See accompanying jpg file: 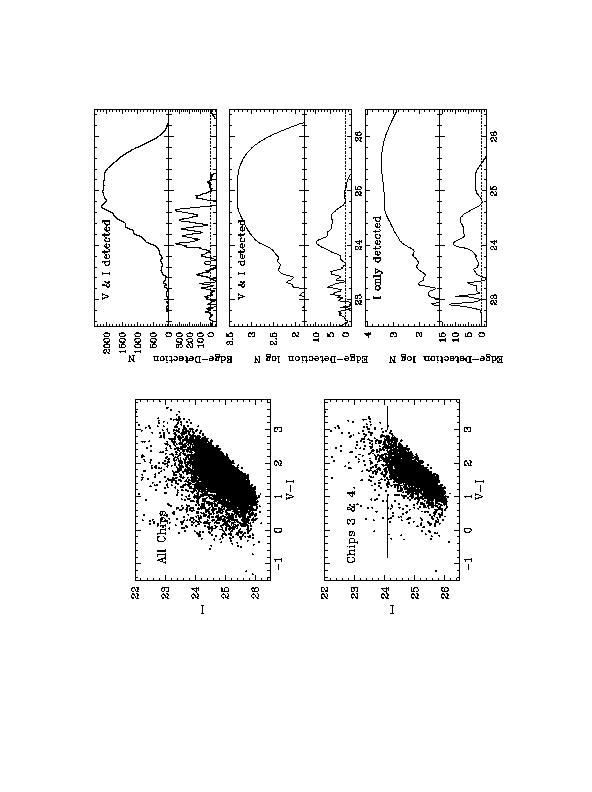}
\label{figure:n3031tip}}

\figcaption[f8.ps]
{NGC~3351: An $I$ vs $(V-I)$ CMD of all stars found in the WFPC2 observations
of NGC~3031 {\it (top left)}, and those found in the halo as indicated by a 
dotted triangle in Figure~1 {\it (bottom left)}.
The linear and logarithmic luminosity functions and corresponding
edge-detection filter response functions are also shown {\it (right)}.
{\bf See accompanying jpg file: 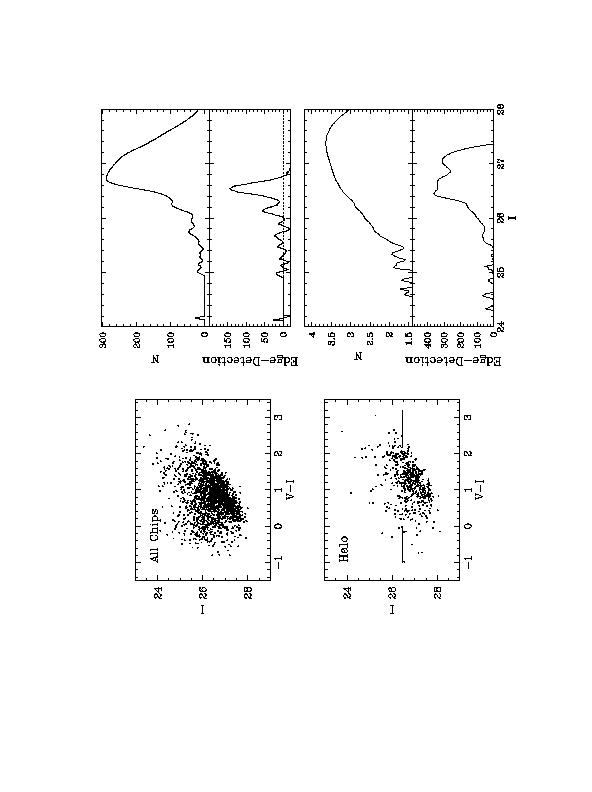}
\label{figure:n3351tip}}

\figcaption[f9.ps]
{NGC~3621:  An $I$ vs $(V-I)$ CMD of all stars found in the WFPC2 observations
of NGC~3621 {\it (top left)}, and those found on the WF2 chip {\it (bottom left)}.
The linear and logarithmic luminosity functions and corresponding
edge-detection filter response functions are also shown {\it (right)}.
{\bf See accompanying jpg file: 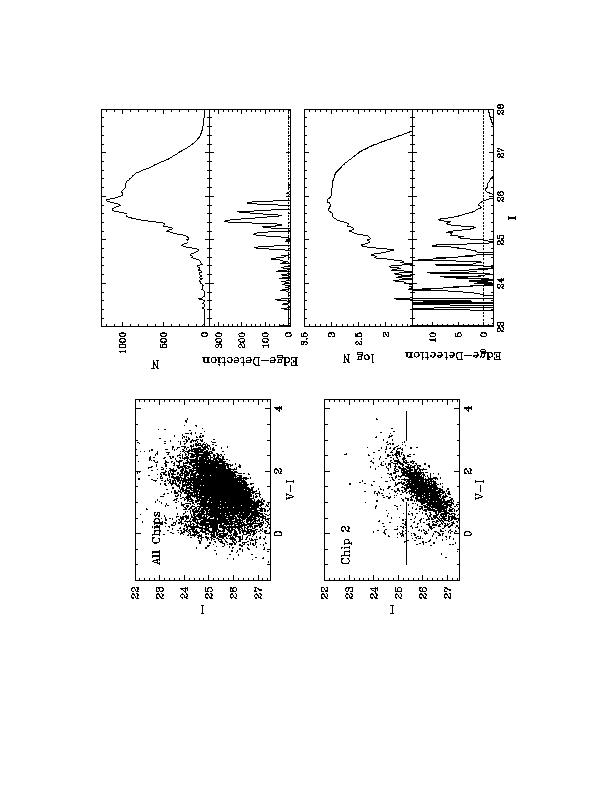}
\label{figure:n3621tip}}

\figcaption[f10.ps]
{NGC~5457: An $I$ vs $(V-I)$ CMD of all stars found in the WFPC2 observations
of NGC~3621 {\it (top left)}, and those found on WF3 and 4 chips {\it (bottom left)}.
The linear and logarithmic luminosity functions and corresponding
edge-detection filter response functions are also shown {\it (right)}.
{\bf See accompanying jpg file: 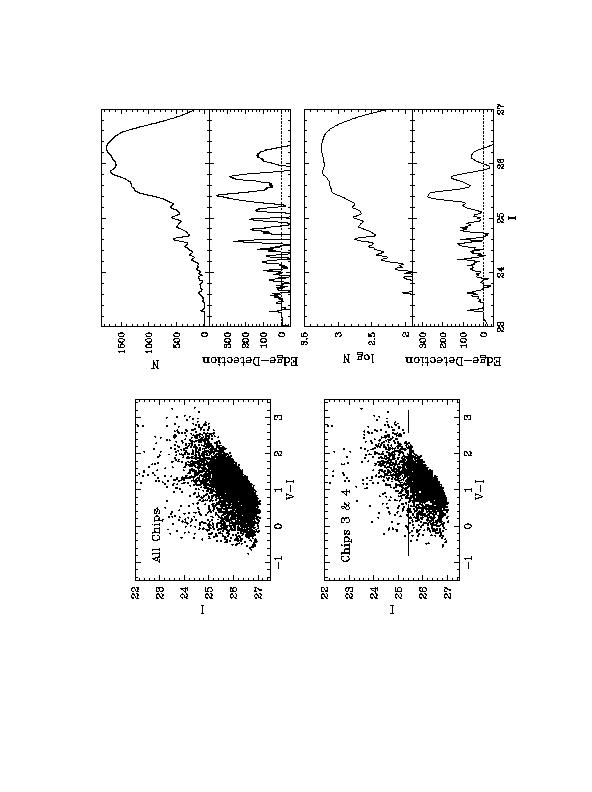}
\label{figure:n5457tip}}

\figcaption[f11.ps]
{In each panel, the I-band luminosity function used for the cross-correlation
method is shown.  The template luminosity function is shifted by the estimated distance,
and shown by the dashed line. 
{\bf See accompanying jpg file: 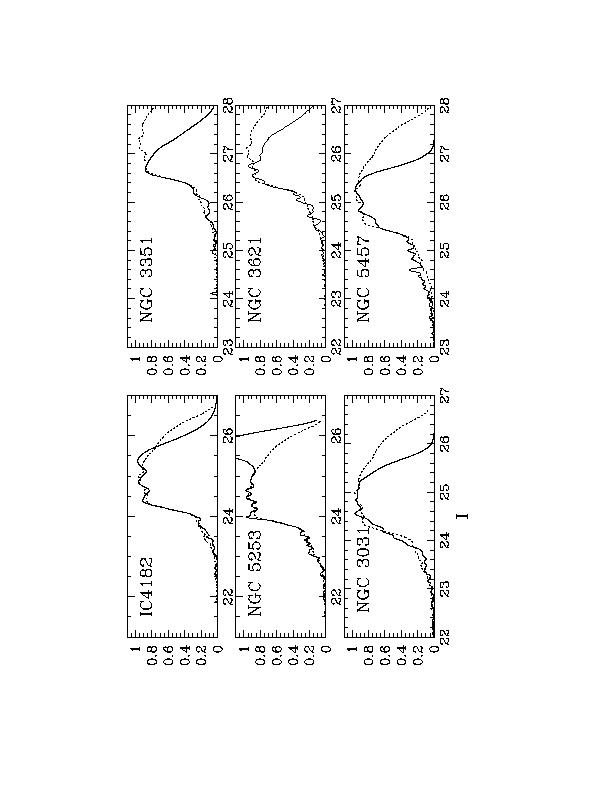}
\label{figure:ccresults}}

\figcaption[f12.ps]
{The difference between TRGB and Cepheid distances for the five different samples discussed in \S 6,
corresponding to the first five samples in Table~5, 
as a function of Cepheid metallicity. Sextans~B does not show in the 2nd and 3rd panels, as it 
deviates by more than 0.5 mag from the fit (see \S6).
{\bf See accompanying jpg file: 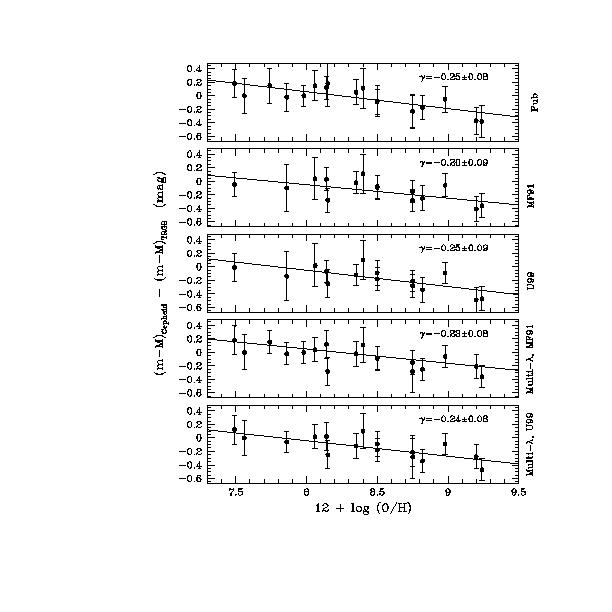}
\label{figure:metaldep}}

\figcaption[f13.ps]
{The difference between TRGB and Cepheid distances as a function of Cepheid metallicity,
with different symbols indicating the different sources of the TRGB distances 
{\it (top)}, and of the Cepheid distances {\it (bottom)}, as indicated in the legend. 
The multi-wavelength sample based on the MF91 scale was used.
{\bf See accompanying jpg file: 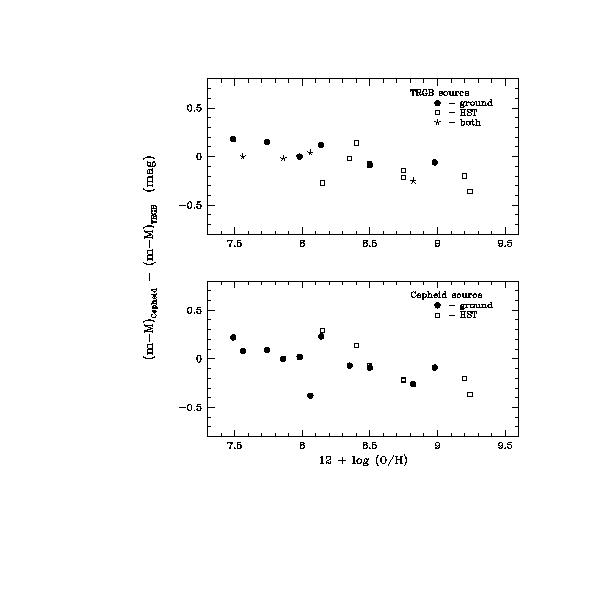}
\label{figure:gbvshst}}

\figcaption[f14.ps]
{Comparison of Pop~I and Pop~II metallicities.
The Pop~II metallicity, estimated by the $V-I$ color of the RGB stars,
affects the magnitude of the TRGB slightly.
This is compared for each galaxy with the Pop~I [O/H] abundances
used to assess the metallicity dependence of Cepheid variable stars
in this paper.  There is at most only a weak correlation
{\bf See accompanying jpg file: 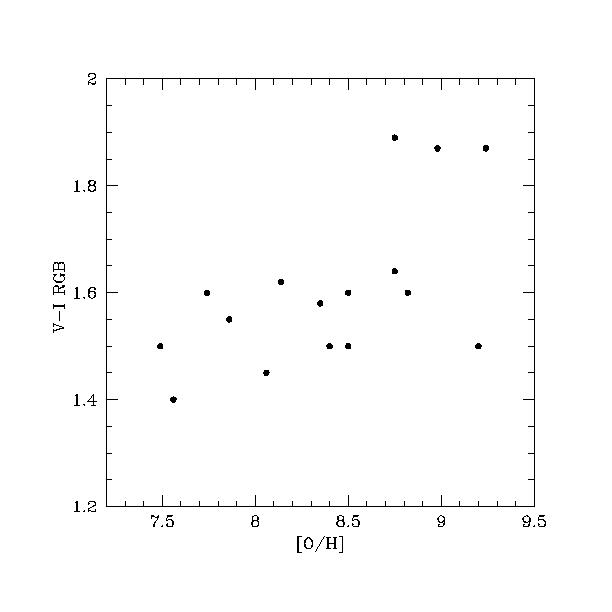}
\label{figure:zcompare}}

\figcaption[f15.ps]
{The difference between TRGB and Cepheid distances as a function of Cepheid metallicity, 
using the abundances based on the ZKH {\it (top)} and $T_e$ scales.
{\bf See accompanying jpg file: 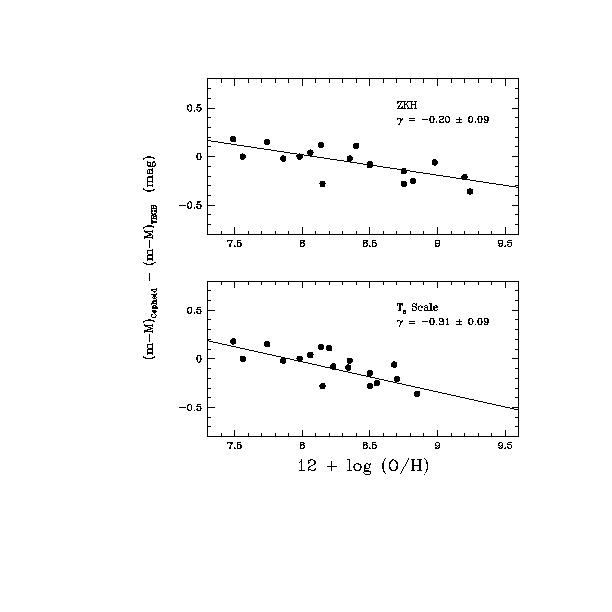}
\label{figure:znew}}

\figcaption[f16.ps]
{Comparison of the metallicity dependence using two independent
TRGB calibrations. The solid circles and line represent the relation based
on the Lee et al. (1993) calibration, while the open circles and dashed line
are those based on  Salaris \& Cassisi 1998. 
{\bf See accompanying jpg file: 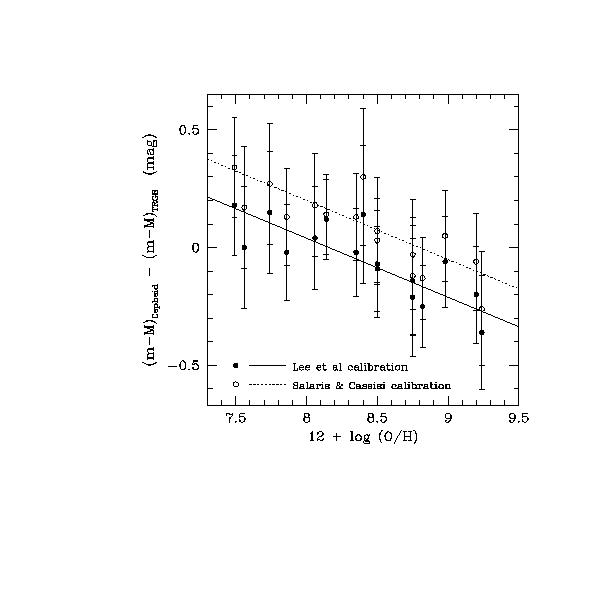}
\label{figure:zpcomp}}

\end{document}